\documentclass[10pt,a4paper]{article}
\usepackage{amsmath, amssymb, theorem, latexsym}

\newtheorem{theorem}{Theorem}[section]
\newtheorem{lemma}[theorem]{Lemma}
\newtheorem{proposition}[theorem]{Proposition}

\newtheorem{hypothesis}[theorem]{Hypothesis}

\newtheorem{remark}[theorem]{Remark}

\newcommand{\eh}{\hfill}\newlength{\sperr}
\newenvironment{proof}{{\settowidth{\sperr}{\rm Proof}
\par\addvspace{0.3cm}\noindent\parbox[t]{1.3\sperr}{\rm P\eh r\eh o\eh o\eh
f\eh.}}}{\nopagebreak\mbox{}\hfill $\blacksquare $\par\addvspace{0.25cm}}
\def \div{{\rm div}}
\def \supp{{\rm supp\,}}
\def \beq{\begin{equation}}
\def \eeq{\end{equation}}

\def\I{\mathcal{I}}

\def \rz { {\mathbb R}}
 
\def\h{\hbar}
\def\Dg {{\mathcal D}}

\def\nz{{\mathbb N}}

\def \rz {{\mathbb R}}
\def \cz {{\mathbb C}}

\def \Tr{{\mathbb{T}\hspace{-1pt}\text{\rm r}\,}}

\def\p{p_{_\mathfrak{z}}}
\newcommand {\ar}{\rightarrow}
\newcommand {\pa}{\partial}

\numberwithin{equation}{section}

\title{Magnetic calculus and semiclassical trace formulas.}
\author{
B. Helffer,\\
Laboratoire de Math\'ematiques, Univ Paris-Sud et CNRS,\\
91 405 Orsay Cedex France,\\
and LEA MathMode\\
and\\
 R. Purice,\\
 Stoilow Institut,\\
and LEA MathMode.
}
\begin{document}
\maketitle

\begin{abstract}
The aim of these notes is to show how the magnetic calculus developed
in \cite{MP, IMP1, IMP2, MPR, LMR} permits to give a new information on the nature of the
coefficients
 of the expansion of the trace of a function of the magnetic
 Schr\"odinger
 operator whose existence was  established in \cite{HR2}.
\end{abstract}

\newpage

\section{Introduction}
Let us consider the magnetic Schr\"odinger operator  on $\mathbb R^d$
 defined by
\beq\label{magn-Lapl}
P^{A}(\h) = \sum_{j=1}^d (\h D_{x_j} -A_j(x))^2 + V(x)\;,
\eeq
where $D_{x_j} :=-i\partial_{x_j}$ and we assume:
\begin{hypothesis}\label{Hyp0}~
\begin{itemize}
\item $\h\in \I\subset]0,+\infty[$, with $\I$ a bounded set having $0$ as accumulation point,
\item $A=(A_1,\dots,A_d)$ with $ A_j \in C^\infty(\mathbb R^d)$,
\item $V\in C^\infty$, $V\geq -C$.
\end{itemize}
\end{hypothesis}

It is known that the operator associated with  $P^A(\h)$ on $C_0^\infty(\mathbb R^d)$
admits
 a unique selfadjoint extension on $L^2(\mathbb R^d)$, which can be
 defined as the Friedrichs extension. We denote by $\widetilde{P^A}(\h)$ this extension.
For any function $g\in C_0^\infty(\mathbb R)$, we can define by the abstract
 functional calculus $g\big(\widetilde{P^A}(\h)\big)$.\\

We shall also make the following assumption concerning the potential function $V$:
\begin{hypothesis}\label{Hyp}
$$\Sigma_V:=\underset{|x|\rightarrow\infty}{\lim\inf}V(x) > \inf V\,.$$
\end{hypothesis}

It is known in this case by Persson's Theorem (see for example
\cite{Ag}) that the spectrum is
discrete in $] -\infty,\Sigma_V[$  and using the max-min principle one shows easily that the spectrum is non empty for $\h$ small enough.
In particular, for $\supp g \subset\subset ]-\infty,
\lim\inf V[ $,  one can consider
$\Tr g \big(\widetilde{P^A}(\h)\big) $.
Our goal is to analyze the expansion of this trace as a power series in $\h$ and
 the dependence of the coefficients on the magnetic field, i.e. the two-form 
$
B := d(\sum_j A_j dx_j)
$. Of course if we have two vector potentials $A$ and $\hat A$, such
that $d A=d \widehat A =B$, we know that there exists $\phi\in
C^{\infty}(\mathbb R^d)$ such that~: $A= \widehat A + d\phi$, and the
conjugation by the 
multiplication operator by $\exp\frac{ i }{\hbar}\phi$ gives a unitary
equivalence between $\widetilde{P^A}(\h)$ and $\widetilde{P^{\widehat
    A}(\h)}$.  Hence $\Tr g \big(\widetilde{P^A}(\h)\big) $ and its
  expansion should depend only on $B$. We would like to investigate
  how it depends effectively on $B$. \\

Our main theorem is the following.

\begin{theorem}\label{Semicl-trace-formula}~\\
Under the previous assumptions on $A$ and $V$ and with
 $H=\widetilde{P^A}(\h)$, there exists a sequence of distributions $T^B_j\in\mathcal{D}^\prime(\mathbb{R})$, $(j\in\mathbb{N})$
 such that for any $ g\in C_0^{\infty}(\mathbb R)$ with $\supp g\subset
]-\infty,\Sigma_V[$ and for any $N\in \mathbb N$, there exist $C_N$ and $h_N$, such that:
\begin{equation}\label{1}
\left|(2\pi \h)^d\Tr\,g(H)\,-\,\sum_{0\leq j\leq
  N}\h^jT^B_j(g)\right|\,\leq\,C_N\h^{N+1}\,,\, \forall \h\in
]0,h_N]\cap \I\,.
\end{equation}
More precisely there exists
$k_j\in \mathbb N$ and universal polynomials  $P_\ell( u_\alpha, v_{\beta,j,k})$ depending on a finite
 number of variables, indexed by $\alpha \in \mathbb N^{2d}$ and
 $\beta\in \mathbb N^d$, such that the distributions:
\begin{equation}
T_j^B (g) = \sum_{0 \leq \ell \leq k_j} \int g ^{(\ell)} (F(x,\xi))
 P_\ell (\pa_{x,\xi}^{\alpha} F(x,\xi), \pa_x^{\beta} B_{jk} (x))\, dx
 d\xi\,,
\end{equation}
where $F(x,\xi)=\xi^2+V(x)$, satisfy \eqref{1}. Finally,  $T_j^B =0$ for $j$ odd. 
\end{theorem}

This theorem was obtained under stronger assumptions in \cite{HR1},
but the main difference with the statement above was 
  that the expression of $T^B_j(g)$ 
  was given in terms of a vector potential $A$ such that $dA=B$. Tricky
 calculations
 permitted after to recover a gauge invariant expression for  the three 
 first terms~:
\begin{equation}
T^B_0(g):=\int_\Xi dx\,d\xi g(F(x,\xi)),\quad T^B_1(g):=0,\quad T^B_2(g):=-\frac{1}{12}\int_\Xi dx\,d\xi g^{\prime\prime}(F(x,\xi))\left[\big(\Delta V\big)(x)+\|B(x)\|^2\right].
\end{equation}
 The approach of \cite{HR1} did not permit  to recover the same
 kind of result for any term of the expansion. On the contrary, we
 will show that, when it can be applied 
  the magnetic calculus permits to give naturally this
 expression. To state the results at the intersection of the domains of validity of
 the two calculi is actually unnecessary. Following essentially 
 arguments presented  
in \cite{HMR} in the case without magnetic potential, we will show 
 how we can use the Agmon exponential decay estimates in order to
 modify the behaviour of $V$ and $A$  at infinity, without changing the asymptotic behavior of $\Tr
 g(\widetilde{P^A(h)} )$, in order to enter simultaneously 
 in Helffer-Robert's
 class and  in the magnetic pseudodifferential calculus of
 \cite{IMP1, IMP2, MP, MPR}.

The paper is organized as follows. In Section \ref{h-calc}, we review
the now standard $\h$-pseudodifferential Weyl calculus and the
attached functional calculus. Section \ref{B-h-calc} is devoted to the presentation of the gauge-invariant
magnetic calculus \cite{IMP1, IMP2, MP, MPR}.
The last section will give the proof of the main theorem.

We shall constantly use the notations $\mathcal{X}\cong\mathbb{R}^d$,
$\Xi:=\mathcal{X}\times\mathcal{X}^\prime$, with $\mathcal{X}^\prime$
the dual of $\mathcal{X}$. The points of $\Xi$ will be denoted as
$X=(x,\xi)$. Recall that $\Xi$ has a canonical symplectic form
$\sigma((x,\xi),(y,\eta)):=\xi(y)-\eta(x)=\underset{1\leq j\leq
  d}{\sum}(\xi_jy_j-\eta_jx_j)$. We denote by 
$C^\infty_{\text{\sf pol}}(\mathcal{Y})$ the space of
$C^\infty$-functions  on the vector space $\mathcal{Y}$ having at
most polynomial growth at infinity together with all their
derivatives. We denote by  $C^\infty_{\text{\sf pol,u}}(\mathcal{Y})$
the  subspace of functions having a unique polynomial upper bound for
all their derivatives.

\paragraph{Acknowledgements} We ackowledge the support of the CNRS Franco-Romanian European Associated Laboratory {\it Math-Mode} and of the University Paris Sud as well as the ANCS Partnership Contract No. 62-056/2008. The second author thanks the University Paris Sud for its kind hospitality during part of the elaboration of this paper.

\section{The $\h$-pseudodifferential Weyl calculus and trace formulas}
\label{h-calc}

 To any Schwartz test function $\phi\in\mathcal{S}(\Xi)$ we associate a bounded linear operator $\mathfrak{Op}_\hbar(\phi)$ on the Hilbert space $\mathcal{H}:=L^2(\mathcal{X})$:
$$
\left[\mathfrak{Op}_\hbar(\phi)u\right](x):=(2\pi\hbar)^{-d}\int_{\mathcal{X}}\int_{\mathcal{X}^\prime}dy\,d\eta\,e^{\frac{i}{\hbar}\eta(x-y)}\phi
\left(\frac{x+y}{2},\eta\right)u(y),\quad\forall u\in\mathcal{S}(\Xi),
$$
for some constant $\hbar>0$. It is easy to prove that $\mathfrak{Op}_\hbar(\phi)\in\mathbb{B}(\mathcal{H})$ and
$$
\|\mathfrak{Op}_\hbar(\phi)\|_{\mathbb{B}(\mathcal{H})}\leq\int_{\mathcal{X}}dx\|[\mathfrak{F}^-\phi](x,\cdot)\|_\infty
$$
where $\mathfrak{F}^-$ denotes the inverse Fourier transform with respect to the second variable. Moreover it is not hard to prove, by
using Schur's Lemma,  that $\mathfrak{Op}_\hbar:\mathcal{S}(\Xi)\rightarrow\mathbb{B}(\mathcal{H})$ extends to an isomorphism of topological vector spaces $\mathfrak{Op}_\hbar:\mathcal{S}^\prime(\Xi)\rightarrow\mathbb{B}\big(\mathcal{S}(\mathcal{X});\mathcal{S}^\prime(\mathcal{X})\big)$. We can transport the operator multiplication from $\mathbb{B}(\mathcal{H})$ back to a non-commutative product on $\mathcal{S}(\Xi)$
$$
\mathfrak{Op}_\hbar(\phi)\mathfrak{Op}_\hbar(\psi)=:\mathfrak{Op}_\hbar(\phi\sharp_\hbar\psi),\quad\forall(\phi,\psi)\in\left[\mathcal{S}(\Xi)\right]^2.
$$
Explicitely we have
\begin{equation}\label{Moyal-prod}
\left(\phi\sharp_{\hbar}\psi\right)(X):=(\pi\hbar)^{-2d}\int_\Xi
dY\int_\Xi dZ\,\exp\left[-\left(2i/\hbar\right)
\sigma(Y,Z)\right]\phi(X-Y)\psi(X-Z).
\end{equation}

One can prove that $$\mathfrak{Op}_\hbar\big[\mathcal{S}(\Xi)\big]=\mathbb{B}\big(\mathcal{S}(\mathcal{X});\mathcal{S}(\mathcal{X})\big)=\mathbb{B}\big(\mathcal{S}^\prime(\mathcal{X});\mathcal{S}^\prime(\mathcal{X})\big)
$$
so that one can consider products of the form $\mathfrak{Op}_\hbar(\Phi)\mathfrak{Op}_\hbar(\phi)$ and $\mathfrak{Op}_\hbar(\phi)\mathfrak{Op}_\hbar(\Phi)$ for $(\Phi,\phi)\in\mathcal{S}^\prime(\Xi)\times\mathcal{S}(\Xi)$ and define the Moyal algebra
$$
\mathfrak{M}(\Xi):=\left\{\Phi\in\mathcal{S}^\prime(\Xi)\mid\Phi\sharp_\hbar\phi\in\mathcal{S}(\Xi),\phi\sharp_\hbar\Phi\in\mathcal{S}(\Xi),\forall\phi\in\mathcal{S}(\Xi)\right\},
$$
that is obviously an algebra for the $\sharp_\hbar$-multiplication and
even a 
$*$-algebra for the anti-involution given by complex conjugation of distributions. 

 The limit $\hbar\rightarrow 0$, that has a rather singular behaviour, should correspond in some sense to the classical algebra of observables that is a commutative algebra. A precise meaning of this limit can be given in the context of strict deformation quantization (see \cite{La}), but is out of our aims in this paper. On the contrary, the asymptotics $\hbar\rightarrow0$, known as the semi-classical asymptotics, and considered in the frame of asymptotic series in $\hbar$ is an important problem and we shall concentrate on some of its aspects when magnetic fields are present.

The Moyal algebra contains many interesting subalgebras, among them the usual H\"{o}rmander symbols 
$$
S^m_1(\mathcal{X}):=\left\{F\in C^\infty(\Xi)\mid\underset{(x,\xi)\in\Xi}{\sup}<\xi>^{-m+|\alpha|}\big|\big(\partial^a_x\partial^\alpha_\xi F\big)(x,\xi)\big|<\infty\right\}.
$$
A symbol $F\in S^m_1(\mathcal{X})$ of strictly positive order $m>0$ satisfying
$$
\exists C>0,\exists R>0\,\text{such that }C<\xi>^m\leq|F(x,\xi)|,\,\forall (x,\xi)\in\Xi\text{ with }|x|\geq R
$$
is called {\it elliptic} and has the property that it exists a positive constant $a\geq0$ such that for any $\mathfrak{z}\in\mathbb{C}\setminus\mathbb{R}\cup(\infty,-a)$ the distribution $\mathfrak{z}+F$ has an inverse $(\mathfrak{z}+F)^-$ with respect to the $\sharp_\hbar$-product and this inverse belongs to the class $S^{-m}_1(\mathcal{X})$. In other words, the operator $\mathfrak{Op}_\hbar(F)$ has a self-adjoint extension with a resolvent that has a symbol of H\"{o}rmander class $S^{-m}_1(\mathcal{X})$. An important problem is how to relate this pseudodifferential calculus (defined by the Moyal product $\sharp_\hbar$) with the usual functional calculus for self-adjoint operators, when  these operators are of the form $\mathfrak{Op}_\hbar(F)$ with $F\in S^m_1(\mathcal{X})$ elliptic ($m>0$).

In dealing with semi-classical problems, the parameter $\hbar$ is no longer constant and it is important to consider asymptotic series in $\hbar$. An essential fact is that the Moyal product $\sharp_\hbar$ has a 'suitable' behaviour with respect to such asymptotic series. More precisely, let us consider the space $S^{(s,m)}_1(\mathcal{X})$ of $\hbar$-symbols of the form
$F:\I\times\Xi\rightarrow\mathbb{C}$ such that 
$$
F(\h)\in C^\infty(\Xi),\ \forall\h\in\I,\quad\underset{\I\times\Xi}{\sup}\ \hbar^{-s}<\xi>^{-m+|\alpha|}\big|\big(\partial^a_x\partial^\alpha_\xi F\big)(\hbar,\xi,x)\big|<\infty.
$$

We refer to  \cite{Rob,Helf-30} for a systematic
discussion and for more general $\hbar$-symbols. In the frame of these
symbol classes, the Moyal product $\sharp_\hbar$ has the following
property  (see  \cite{Rob,Helf-30}):
$$
\forall(F,G)\in S^{(s_1,m_1)}_1(\mathcal{X})\times S^{(s_2,m_2)}_1(\mathcal{X}),\quad F\sharp_\hbar G\in S^{(s_1+s_2,m_1+m_2)}_1(\mathcal{X})\text{ and}
$$
$$
F\sharp_\hbar G-FG\in S^{(s_1+s_2+1,m_1+m_2-1)}_1(\mathcal{X}).
$$
We shall usually work with elements $F\in S^{(s,m)}_1(\mathcal{X})$ having an asymptotic expansion  of the form:
\begin{equation}\label{h-dev}
F(\hbar,\xi,x)\sim\hbar^s\sum_{k\in\mathbb{N}}\hbar^kF_k(x,\xi),\quad\text{with }F_k\in S^{m-k}_1(\mathcal{X}).
\end{equation}
The symbol $\hbar^sF_0\in S^{(s,m)}_1(\mathcal{X})$ is then called {\it the principal symbol of $F$}.

What is important here is to find a class of functions (actually
essentially $C_0^\infty$) such that $g(F)$ is a nice
pseudo\-differential operator with simple
rules
 of computation for the principal symbol.
We are starting from the general Dynkin-Helffer-Sj\"{o}strand formula (see \cite{DiSj})
\begin{equation}\label{DHS} g(P) = -\pi^{-1} \lim_{\epsilon\ar 0^+} \int \!\!\int_{|\mu|\geq \epsilon}
\frac{\pa {\tilde g}}{\pa {\bar z}}(\lambda,\mu)\; (\lambda+i\mu-P)^{-1} d\lambda\,d\mu \,,
\end{equation}
which is true for any selfadjoint operator $P$ and any $g$ in
$C_0^\infty(\rz)$.\\
Here the function  $(\lambda,\mu)\mapsto {\tilde g}(\lambda,\mu)$ is a compactly supported, almost
analytic extension of $g$ to  $\cz$. This means that ${\tilde g} = g$
on $\rz$ and that for any $N\in \nz$ there exists a constant $C_N$
such that  $|\frac{\pa {\tilde g}}{\pa {\bar z}}(\lambda,\mu)|\leq
C_N |\mu|^N\;$.\\ The main result 
due to Helffer-Robert (see also \cite{DiSj} and references therein)  is that, for $P=\mathfrak{Op}_\hbar(F)$ a self-adjoint operator with a $h$-symbol $F$ real and semibounded from below and having an asymptotic expansion  as above (\ref{h-dev}) with $s=0$ and  $g$ in
 $C_0^\infty(\rz)$, the operator  $g(P)$ 
 is a $\hbar$-pseudodifferential operator of the form $\mathfrak{Op}_\hbar(\widetilde{g_\hbar}(F))$, whose Weyl symbol $\widetilde{g_\hbar}(F)(\hbar,\xi,x)$
 admits a formal asymptotic expansion in $\hbar$
\begin{equation}\label{2.4}
\widetilde{g_\hbar}(F)(\hbar,\xi,x)\sim \sum_{k\geq 0} h^k g_{k}(F) (x,\xi)\,,
\end{equation}
with
\begin{equation}
\begin{array}{ll}
g_{0}(F) &= g(F_0)\,,\\
g_{1}(F)&=F_1 \cdot g'(F_0)\,,\\
g_{k}(F)&= \sum_{l=1}^{2k-1}d_{k,l} g^{(l)} (F_0)\,,
\;\forall k\geq 2\,,
\end{array}
\end{equation}
where the $d_{k,l}$ are universal polynomial functions of the symbols $\pa_x^\alpha\pa_\xi^\beta F_\ell$\,, with $|\alpha| + |\beta| +\ell \leq k$\,.
For $k=2$, one has
\begin{equation}
d_{2,1}=F_2,\quad d_{2,2}=p_{2,2}+(1/2)F_1^2,\quad d_{2,3}=p_{2,3}
\end{equation}
with 
\begin{equation}
p_{2,2}(x,\xi)=\frac 18 \sum_{j,k} \left( \frac{\pa^2F_0}{\pa_{x_j}\pa_{\xi_k}}\,
\frac{\pa^2F_0}{\pa {x_k}\pa {\xi_j}}\,-\,
\frac{\pa^2F_0}{\pa {x_j}\pa {x_k}}\,\frac{\pa^2F_0}{\pa
{\xi_j}\pa {\xi_k}}             \right)
\end{equation}
\begin{equation}\label{defp23}
p_{2,3}(x,\xi) = \frac{1}{24}\sum_{j,k}
 \left( 2 \pa^2_{x_k\xi_j}F_0\cdot \pa_{x_j}F_0\cdot \pa_{\xi_k}F_0
 - \pa^2_{x_jx_k}F_0\cdot \pa_{\xi_j}F_0\cdot \pa_{\xi_k}F_0
 - \pa^2_{\xi_j\xi_k}F_0\cdot \pa_{x_j}F_0\cdot
 \pa_{x_k}F_0\right)\,.
\end{equation}

\noindent The main  point in the proof is that one can construct for
$\Im z \neq 0$ 
a parametrix (= approximate inverse) for
$(P-z)$  with a nice control as $\Im z$ tends to $0$.
The constants controlling the estimates on the symbols are exploding
as $\Im z\ar 0$ but the choice of the almost analytic extension of $f$
 absorbs any negative power of $|\Im z|$.

\noindent As a consequence, one gets that for $\hbar$ small enough, if  for  some  interval
$I$ and some $\epsilon_0>0$,  
\begin{equation}\label{H4}
\quad F_0^{-1} ( I + [-\epsilon_0,\epsilon_0]) 
{\rm\;is \; compact},
\end{equation}
 then the spectrum of $\mathfrak{Op}_\hbar(F_0)$ is discrete in $I$.
In particular, one gets that, if $F_0 (x,\xi)\ar +\infty$ as $|x|+|\xi|
\ar +\infty$, then the spectrum of $\mathfrak{Op}_\hbar(F_0)$ is discrete ($\mathfrak{Op}_\hbar(F_0)$ has
compact resolvent). In fact one gets more precisely the
following theorem (due to Helffer-Robert).
\begin{theorem}:\\
Let $P=\mathfrak{Op}_\hbar(F)$ be a self-adjoint operator with a $\h$-symbol $F$ real and semibounded from below, having an asymptotic expansion of the form (\ref{h-dev}) and also satisfying (\ref{H4}) with $I=[E_1,E_2]$,
then,  for  any $g$ in  $C_0^\infty(]E_1,E_2[)$, we have the following
expansion in powers of $\hbar$~:
\begin{equation} {\rm Tr\;} [g\big(\mathfrak{Op}_\hbar(F)\big)] \sim 
 (2\pi\hbar)^{-d} \sum_{j\geq 0}\hbar^j
\;T_j(g)\,,
\end{equation} 
where $g
\mapsto T_j(g)$ are distributions in $\Dg'(]E_1,E_2[)$.\\ In
particular we have, when $F_1=F_2=0$,
 \begin{equation}
\begin{array}{ll}
T_0(g) &= \int \!\!\int g(F_0 (x,\xi))\;dx\, d\xi\,,\\
T_1(g) &= 0\,,\\
T_2(g) 
&  = - \frac{1}{24} \int \!\!\int  g''(F_0(x,\xi)) 
\sum_{j,k} \left( \frac{\pa^2F_0}{\pa {\xi_j}\pa {\xi_k}}\,
\frac{\pa^2F_0}{\pa {x_j}\pa {x_k}}\,-\,
\frac{\pa^2F_0}{\pa {x_j}\pa {\xi_k}}\,\frac{\pa^2F_0}{\pa
{\xi_j}\pa {x_k}}\right)\,dxd\xi\,.
\end{array}
\end{equation}
\end{theorem}

This theorem is obtained by integration of the symbol of $g\big(\mathfrak{Op}_\hbar(F)\big)$ given in \eqref{2.4}, 
because we have the needed regularity so that the trace of a trace-class
operator $\mathfrak{Op}_\hbar(F)$ is given by the integral of the symbol
$F$ 
over $\Xi$. According to the definition
of the Weyl quantization, 
the distribution kernel is  given by the oscillatory integral:
\begin{equation}\label{noyau}
K(x,y;h) = (2\pi \h)^{-d}\int_{\mathcal{X}} \exp \left(\frac{i}{\h} (x-y)\cdot \xi\right)
\;\;F\left(\hbar,\xi,\frac{x+y}{2}\right)\, d\xi\;,
\end{equation}
and the trace of $\mathfrak{Op}_\hbar(F)$ is the integral over $\mathcal{X}$ of the restriction to the
diagonal of the distribution kernel: $K(x,x)=(2\pi \h)^{-d}\int_{\mathcal{X}} 
\;\;F(\hbar,\xi,x)\, d\xi$.

Of course, one could
think of using  the theorem 
 with $g$ the characteristic function of an interval,
 in order to get for example, the
behavior of the counting function attached to this interval. This is of course not directly
possible
 and this will be only obtained through Tauberian theorems
 (\cite{Ho2}, \cite{Ho} and \cite{Iv})  and at the
 price of additional errors. 
Let us however remark that, if the function $g$ is not regular, then
the length of the expansion depends on the regularity of $g$. So it
will not be surprising that, by looking at the Riesz means:
$
g_{s,E}(t)\,:=\,\max\left\{0,(E-t)\right\}^s
$
(for some $s\geq0$ and $E\in(E_1,E_2)$), we shall get a better expansion when $s$ is large.

Under some assumptions on $A$ and $V$, including the condition $\div
A=0$, one can show that $P_A(\h)$ is an $\h$-pseudodifferential operator
whose total Weyl-symbol is  $F(x,\xi)=(\xi -A)^2 + V$ in some class
of \cite{HR1}. More prcsely, we have to assume that for any $\alpha \in
\mathbb N^d $, we have
\beq
|\pa_x^\alpha V(x)|\leq C_\alpha (V(x)+C+1)\,,
\eeq
and, for $j=1\,\dots, d$, the following non gauge covariant conditions:
\beq
|\pa_x^\alpha A_j(x)|\leq C_\alpha (V(x)+C+1)^\frac 12\,.
\eeq

\section{The $\h$-magnetic quantization}\label{B-h-calc}

\subsection{Results for fixed $\h$}

We consider a magnetic field described by a bounded smooth closed $2$-form $B$ on $\mathcal{X}\equiv\mathbb{R}^d$ and the associated modified symplectic form on $\Xi$
$$
\sigma^B_{(x,\xi)}((y,\eta),(\zeta,z)):=\eta(z)-\zeta(y)+B_x(y,z),
$$
that may be used to define the classical Hamiltonian system in the given magnetic field. For the quantum description we shall also consider the Hilbert space $\mathcal{H}=L^2(\mathcal{X})$, we shall choose a smooth vector potential $A$, i.e. a 1-form satisfying the equality $B=dA$ and we shall define the following gauge covariant representation, for all  $ \phi\in\mathcal{S}(\Xi)$ and all $u\in\mathcal{S}(\Xi)$,
$$
\left[\mathfrak{Op}^A_\hbar(\phi)u\right](x):(2\pi\hbar)^{-d}\int_{\mathcal{X}}\int_{\mathcal{X}^\prime}dy\,d\eta\,e^{\frac{i}{\hbar}\eta(x-y)}e^{-\frac{i}{\hbar}\int_{[x,y]}A}\phi\left(\frac{x+y}{2},\eta\right)u(y)\,,
$$
 where $\int_{[x,y]}A$ denotes the integration of the $1$-form $A$ along the oriented segment $[x,y]$.
This gauge covariant 'magnetic quantization' allows to define a 'magnetic' Moyal product $\sharp^B_\hbar$
\begin{equation}\label{B-Moyal-prod}
\begin{array}{ll}
\left(\phi\sharp^B_{\hbar}\psi\right)(X)&:=(\pi\hbar)^{-2d}\int_\Xi
dY\int_\Xi dZ\,\exp\left[-\left(2i/\hbar\right)
\sigma^B_x(Y,Z)\right]\phi(X-Y)\psi(X-Z)\\
& =(\pi\hbar)^{-2d}\int_\Xi
dY\int_\Xi dZ\,e^{-\left(2i/\hbar\right)
\sigma(Y,Z)}e^{-\left(i/\hbar\right)\theta^B(x,y,z)}\phi(X-Y)\psi(X-Z)\\
&
=(\pi\hbar)^{-2d}\int_\Xi
dY\int_\Xi dZ\,e^{-\left(2i/\hbar\right)
\sigma(X-Y,X-Z)}e^{-\left(i/\hbar\right)\widetilde{\theta^B}(x,y,z)}\phi(Y)\psi(Z).
\end{array}
\end{equation}
$$
\theta^B(x,y,z):=\int_{<x-y-z,x+y-z,x-y+z>}\hspace{-2cm}B\hspace{2cm},\quad\widetilde{\theta^B}(x,y,z):=\int_{<x-y+z,y-z+x,z-x+y>}\hspace{-2cm}B\hspace{2cm}.
$$
 Here $<x-y-z,x+y-z,x-y+z>$ denotes the triangle defined by the three points
 $x-y-z$,$x+y-z$, and $x-y+z$, with the usual trigonometric orientation and the integrals of $B$ denote the integrals of the two form on the given oriented triangle.
Associated with  this product we can define a 'magnetic' Moyal algebra for the magnetic field $B$:
$$
\mathfrak{M}^B(\Xi):=\left\{\Phi\in\mathcal{S}^\prime(\Xi)\mid\Phi\sharp^B_\hbar\phi\in\mathcal{S}(\Xi),\phi\sharp^B_\hbar\Phi\in\mathcal{S}(\Xi),\forall\phi\in\mathcal{S}(\Xi)\right\},
$$
This 'magnetic' Moyal calculus preserves a large number of the nice features of the usual Moyal calculus and we shall recall some of them that are useful for our analysis of semiclassical trace formulas.
\begin{proposition}\label{P1} (Propositions 3.5 and 3.10 in \cite{MP})\\
For any magnetic field $B$ with components of class $C^\infty_{\text{\sf pol}}(\mathcal{X})$, one can find a vector potential $A$ with components also of class  $C^\infty_{\text{\sf pol}}(\mathcal{X})$ and then the application $\mathfrak{Op}^A_\hbar:\mathcal{S}(\Xi)\rightarrow\mathbb{B}(L^2(\mathcal{X}))$ extends to an isomorphism of vector spaces $\mathfrak{Op}^A_\hbar:\mathcal{S}^\prime(\Xi)\rightarrow\mathbb{B}(\mathcal{S}(\mathcal{X});\mathcal{S}^\prime(\mathcal{X}))$. The above isomorphism has a restriction $\mathfrak{Op}^A_\hbar:L^2(\Xi)\rightarrow\mathbb{B}_2(L^2(\mathcal{X}))$ that is unitary (here $\mathbb{B}_2(L^2(\mathcal{X}))$ is the algebra of Hilbert-Schmidt operators on $L^2(\mathcal{X})$).
\end{proposition}
\begin{proposition}\label{P2} (Proposition 4.23 in \cite{MP} and Lemma 1.2 in \cite{IMP1})~\\
For any magnetic field $B$ with components of class $C^\infty_{\text{\sf pol}}(\mathcal{X})$, we have the following inclusions:
$$
C^\infty_{\text{\sf pol,u}}(\Xi)\subset\mathfrak{M}^B(\Xi);\qquad S^m_1(\mathcal{X})\subset\mathfrak{M}^B(\Xi).
$$
\end{proposition}
\begin{proposition}\label{P3} (Theorem 2.11 in \cite{LMR})~\\
For any magnetic field $B$ with components of class $BC^\infty(\mathcal{X})$ the 'magnetic' Moyal product defines a continuous map
$$
S^{m_1}_1(\mathcal{X})\times S^{m_2}_1(\mathcal{X})\ni(F,G)\mapsto F\sharp^B_\hbar G\in S^{m_1+m_2}_1(\mathcal{X}),
$$
and for any $N\in\mathbb{N}$ there is a canonical expansion 
$$
F\sharp^B_\hbar G=\sum_{j=0}^{N-1}H_j+R_N,\quad\text{with }H_j\in S^{m_1+m_2-j}_1(\mathcal{X}),R_N\in S^{m_1+m_2-N}_1(\mathcal{X})
$$
in which $H_0=F\cdot G$.
\end{proposition}
\begin{proposition}\label{P4} (Theorem 3.1 in \cite{IMP1})~\\
For any magnetic field $B$ with components of class $BC^\infty(\mathcal{X})$ we have that for any associated vector potential $A$,
$$
\mathfrak{Op}^A_\hbar\big[S^0_1(\mathcal{X})\big]\subset\mathbb{B}(L^2(\mathcal{X}))
$$
and there exist two positive constants $c,p$ depending only on the dimension $d$ of $\mathcal{X}$ such that
$$
\left\|\mathfrak{Op}^A_\hbar(F)\right\|_{\mathbb{B}(L^2(\mathcal{X}))}\leq c\underset{|a|\leq p}{\sup}\big(\underset{|\alpha|\leq p}{\sup}\big(\underset{(x,\xi)\in\Xi}{\sup}<\xi>^{|\alpha|}\left|\big(\partial^a_x\partial^\alpha_\xi F\big)(x,\xi)\right|\big)\big).
$$
\end{proposition}
\begin{proposition}\label{P5} (Theorems 4.1 and 4.3 in \cite{IMP1} and Proposition 6.31 in \cite{IMP2})~\\
Suppose the magnetic field $B$ is of class $BC^\infty(\mathcal{X})$; then
\begin{itemize}
 \item if $F\in S^0_1(\mathcal{X})$ is a real function, $\mathfrak{Op}^A_\hbar(F)$ is a bounded self-adjoint operator on $L^2(\mathcal{X})$ for any vector potential $A$ of $B$; then the resolvent of $\mathfrak{Op}^A_\hbar(F)$ has a 'magnetic' symbol of class $S^0_1(\mathcal{X})$;
\item if $F\in S^m_1(\mathcal{X})$ is a real elliptic symbol with $m>0$, then $\mathfrak{Op}^A_\hbar(F)$ has a self-adjoint extension in $L^2(\mathcal{X})$  for any vector potential $A$ of $B$ and the resolvent has a 'magnetic' symbol of class $S^{-m}_1(\mathcal{X})$; if we choose $A$ with components of class $C^\infty_{\text{\sf pol}}(\mathcal{X})$ then $\mathfrak{Op}^A_\hbar(F)$ is essentially self-adjoint on $\mathcal{S}(\mathcal{X})$ and its self-adjoint extension has as domain the 'magnetic' Sobolev space:
$$
\mathcal{H}^A_m(\mathcal{X}):=\left\{u\in L^2(\mathcal{X})\mid\mathfrak{Op}^A_\hbar(p_m)u\in L^2(\mathcal{X}),\text{ where }p_m(x,\xi):=<\xi>^m\right\};
$$
\item if $F\in S^m_1(\mathcal{X})$, with $m\in\mathbb{R}$, satisfies $\Re F(x,\xi)\geq C|\xi|^m$ for $|\xi|\geq R$, with some strictly positive constants $C$ and $R$, then for any $r\in\mathbb{R}$ there exist two positive constants $C_0$ and $C_1$ such that for any $u\in\mathcal{H}^A_\infty(\mathcal{X})$ we have
$$
\Re<u,\mathfrak{Op}^A(F)u>_{L^2(\mathcal{X}}\geq C_0\|u\|_{\mathcal{H}^A_{m/2}(\mathcal{X})}-C_1\|u\|_{\mathcal{H}^A_{s}(\mathcal{X})\,.}
$$
\end{itemize}
\end{proposition}
\begin{proposition}\label{P6} (Proposition 6.33 in \cite{IMP2})~\\
Suppose the magnetic field $B$ has components of class $BC^\infty(\mathcal{X})$ and $\Phi\in C^\infty_0(\mathbb{R})$, then,
 for any real $F\in S^m_1(\mathcal{X})$ with $m\geq0$, 
$F$ elliptic for $m>0$, and for any $A$ such that $dA=B$, the operator $\Phi\big(\mathfrak{Op}^A_\hbar(F)\big)$, defined by the 
functional calculus for  self-adjoint operators  has a 'magnetic' symbol of class $S^{-m}_1(\mathcal{X})$.
\end{proposition}

\subsection{Semiclassical results}

Let us consider now the dependence on $\hbar\in\I$. We shall come back to (\ref{B-Moyal-prod}) and analyze the $\hbar$-dependence of this product:
$$
\left(\phi\sharp^B_{\hbar}\psi\right)(X)=(\pi\hbar)^{-2d}\int_\Xi
dY\int_\Xi dZ\,e^{-\left(2i/\hbar\right)
\sigma(Y,Z)}e^{-\left(i/\hbar\right)\theta^B(x,y,z)}\phi(X-Y)\psi(X-Z)
$$
with
$$
\theta^B(x,y,z):=\int_{<x-y-z,x+y-z,x-y+z>}\hspace{-2cm}B\hspace{2cm}=4\sum_{jk}y_jz_k\int_0^1\hspace{-0.2cm}ds\int_0^{1-s}\hspace{-0.5cm}dt\,B_{jk}\big(x+(2s-1)y+(2t-1)z\big).
$$
Let us make the change of variables $(y,z)\mapsto(\h y,\h z)$ in order to obtain
\begin{equation}\label{B-Moyal-prod-1}
(\phi\sharp^B_\h \psi)(X)\end{equation}
$$
=\pi^{-2d}\int_{\Xi}dY\hspace{-0.2cm}\int_{\Xi}dZ \ e^{-2i\sigma(Y,Z)}e^{-(i\h)\theta^B_\hbar(x,y,z)}\phi(x-\h y,\xi-\eta)\psi(x-\h z,\xi-\zeta),
$$
with
$$
\theta^B_\hbar(x,y,z)=4\sum_{jk}y_jz_k\int_0^1\hspace{-0.2cm}ds\int_0^{1-s}\hspace{-0.5cm}dt\,B_{jk}\big(x+(2s-1)\h y+(2t-1)\h z\big).
$$
We notice that we can now obtain an asymptotic expansion  of the $\sharp^B_\hbar$-product with respect to $\hbar$ by using the Taylor formulas for $\phi$, $\psi$, like in the non-magnetic case, and for the exponential $e^{-(i\h)\theta^B_\hbar(x,y,z)}$ and also for $B$ in the expression of $\theta^B_\hbar(x,y,z)$:
$$
\phi(x-\h y,\xi-\eta)=\sum_{0\leq\nu\leq N}\frac{(-\h)^\nu}{\nu!}\sum_{|\alpha|=\nu}\frac{\nu!}{\alpha!}y^\alpha\big(\partial_x^\alpha \phi\big)(x,\xi-\eta)+\mathfrak{R}_{\phi,N},
$$
$$
\psi(x-\h z,\xi-\zeta)=\sum_{0\leq\mu\leq N}\frac{(-\h)^\mu}{\mu!}\sum_{|\beta|=\mu}\frac{\mu!}{\beta!}z^\beta\big(\partial_x^\beta \psi\big)(x,\xi-\zeta)+\mathfrak{R}_{\psi,N},
$$
$$
e^{-(i\h)\theta^B_\hbar(x,y,z)}=\sum_{0\leq\rho\leq N}\frac{(-i\h)^\rho}{\rho!}\left[\theta^B_\hbar(x,y,z)\right]^\rho+\mathfrak{R}_{B,N},
$$
$$
\theta^B_\hbar(x,y,z)=\sum_{0\leq\lambda\leq N}\frac{\h^\lambda}{\lambda!}\left[\sum_{|\gamma|=\lambda}\left(\sum_{jk}y_jz_k\left(\partial^\gamma B_{jk}\right)(x)\right)\frac{\lambda!}{\gamma!}\int_{-1}^1\hspace{-0.2cm}ds\int_{-1}^{-s}\hspace{-0.5cm}dt\big(sy+tz\big)^\gamma\right]+\mathfrak{r}_{B,N}$$
$$
=\sum_{0\leq\lambda\leq N}\frac{\h^\lambda}{\lambda!}\left[\sum_{|\gamma|=\lambda}\sum_{\delta\leq\gamma}\boldsymbol{T}^\gamma_\delta\ y^\delta z^{\gamma-\delta}\left(\sum_{jk}y_jz_k\left(\partial^\gamma B_{jk}\right)(x)\right)\right]+\mathfrak{r}_{B,N},
$$
where for any $N\geq 1$ in $\mathbb{N}$ we have
$$
\mathfrak{R}_{\phi,N}(\h,x,y,\xi,\eta)=(-\h)^{N+1}\sum_{|\alpha|=N+1}\frac{y^\alpha}{\alpha!}\int_0^1\big(\partial^\alpha_x \phi\big)(x-u\h y,\xi-\eta)du
=\sum_{|\alpha|=N+1}y^\alpha\widetilde{\mathfrak{R}}_{\phi,N,\alpha}(\h,x,y,\xi,\eta),
$$
$$
\mathfrak{R}_{\psi,N}(\h,x,z,\xi,\zeta)=(-\h)^{N+1}\sum_{|\beta|=N+1}\frac{z^\beta}{\beta!}\int_0^1\big(\partial^\alpha_x \psi\big)(x-u\h z,\xi-\zeta)du=\sum_{|\beta|=N+1}z^\beta\widetilde{\mathfrak{R}}_{\phi,N,\beta}(\h,x,z,\xi,\zeta),
$$
$$
\mathfrak{R}_{B,N}(\h,x,y,z)=(-i\h)^{N+1}\left[\theta^B_\h(x,y,z))\right]^{N+1}\int_0^1du_1\ldots\int_0^{u_N}e^{-iu_{N+1}\h \theta^B_\h(x,y,z))}du_{N+1}$$
$$
=\h^{N+1}\chi^B_{N+1}(\h,x,y,z)\left[\sum_{0\leq\lambda\leq N}\frac{\h^\lambda}{\lambda!}\left[\sum_{|\gamma|=\lambda}\sum_{\delta\leq\gamma}\boldsymbol{T}^\gamma_\delta\ y^\delta z^{\gamma-\delta}\left(\sum_{jk}y_jz_k\left(\partial^\gamma B_{jk}\right)(x)\right)\right]+\mathfrak{r}_{B,N}\right]^{N+1}
$$
$$
\mathfrak{r}_{B,N}(\h,x,y,z)=\h^{N+1}\hspace{-0.4cm}\sum_{|\gamma|=N+1}\hspace{-0.2cm}(\gamma!)^{-1}\hspace{-0.2cm}\int_{-1}^1\hspace{-0.2cm}ds\int_{-1}^{-s}\hspace{-0.5cm}dt\int_0^1\hspace{-0.2cm}du\sum_{jk}y_jz_k\left(\partial^\gamma B_{jk}\right)\big(x+\h u(sy+tz)\big)(sy+tz)^\gamma$$
$$
=\h^{N+1}\sum_{jk}y_jz_k\hspace{-0.4cm}\sum_{|\gamma|=N+1}\hspace{-0.2cm}(\gamma!)^{-1}F^B_{\gamma,j,k}(\h,x,y,z)\sum_{\delta\leq\gamma}\boldsymbol{T}^\gamma_\delta\ y^\delta z^{\gamma-\delta}.
$$
Let us observe first that the powers $y^\alpha$ and $z^\beta$ appearing in the Taylor expansions of $\phi$ and $\psi$ are dealt by integration by parts, using the 'oscillatory' exponential $e^{-2i\sigma(Y,Z)}$ and transformed in $\zeta$-derivatives of $\psi$ and respectively in $\eta$-derivatives of $\phi$. A similar procedure may be used to get rid of the factors $y^\delta z^\gamma$ appearing in the expansions of $\theta^B_\hbar(x,y,z)^\rho$; in fact we have
$$
y^\alpha e^{-2i\sigma(Y,Z)}=(-i/2)^{|\alpha|}\partial_\zeta^\alpha e^{-2i\sigma(Y,Z)};\qquad z^\beta e^{-2i\sigma(Y,Z)}=(i/2)^{|\beta|}\partial_\eta^\beta e^{-2i\sigma(Y,Z)}.
$$

Thus, denoting by
\begin{equation}\label{rest-T}
\mathfrak{T}^B_\lambda(x):=\sum_{\begin{array}{c}\scriptstyle{|\gamma|=\lambda}\\\scriptstyle{\delta\leq\gamma}\end{array}}\sum_{jk}(-1)^{\lambda-|\delta|}\boldsymbol{T}^\gamma_\delta\left(\partial^\gamma B_{jk}\right)(x)\partial_\zeta^\delta\partial_{\zeta_j} \partial_\eta^{\gamma-\delta}\partial_{\eta_k},
\end{equation}
one has
$$
(\phi\sharp^B_\h \psi)(X)=\pi^{-2d}\int_{\Xi}dY\hspace{-0.2cm}\int_{\Xi}dZ \ e^{-2i\sigma(Y,Z)}\times
$$
$$
\times\left\{\sum_{0\leq\rho\leq N}\frac{(-i\h)^\rho}{\rho!}\left[\sum_{0\leq\lambda\leq N}\frac{\h^\lambda}{4\lambda!}\left(\frac{i}{2}\right)^\lambda\mathfrak{T}^B_\lambda(x)+\mathfrak{r}_{B,N}\right]^\rho+\mathfrak{R}_{B,N}\right\}\times
$$
$$
\times\left[\sum_{0\leq\nu\leq N}\frac{(-\h)^\nu}{\nu!}\sum_{|\alpha|=\nu}\frac{\nu!}{\alpha!}\left((i/2)\partial_\zeta\right)^\alpha\big(\partial_x^\alpha \phi\big)(x,\xi-\eta)+\mathfrak{R}_{\phi,N}\right]\times
$$
$$
\times\left[\sum_{0\leq\mu\leq N}\frac{(-\h)^\mu}{\mu!}\sum_{|\beta|=\mu}\frac{\mu!}{\beta!}\left((-i/2)\partial_\eta\right)^\beta\big(\partial_x^\beta \psi\big)(x,\xi-\zeta)+\mathfrak{R}_{\psi,N}\right]$$
\begin{equation}\label{main-dev}
=(2\pi)^{-2d}\left\{\sum_{0\leq\rho\leq N}\frac{(-i\h)^\rho}{\rho!}\left[\sum_{0\leq\lambda\leq N}\frac{\h^\lambda}{4\lambda!}\left(\frac{i}{2}\right)^\lambda\mathfrak{T}^B_\lambda(x)\right]^\rho\right\}\times
\end{equation}
$$
\times\left[\sum_{\begin{array}{c}\scriptstyle{0\leq\nu\leq N}\\\scriptstyle{0\leq\mu\leq N}\end{array}}\frac{(i)^{\nu+\mu}(-1)^\mu(\h)^{\nu+\mu}}{2^{\nu+\mu}\nu!\mu!}\sum_{\begin{array}{c}\scriptstyle{|\alpha|=\nu}\\\scriptstyle{|\beta|=\mu}\end{array}}\frac{\nu!}{\alpha!}\frac{\mu!}{\beta!}\big(\partial_\xi^\beta\partial_x^\alpha \phi\big)(x,\xi)\big(\partial_\xi^\alpha\partial_x^\beta \psi\big)(x,\xi)\right]+
$$
$$
+\pi^{-2d}\int_{\Xi}dY\hspace{-0.2cm}\int_{\Xi}dZ \ e^{-2i\sigma(Y,Z)}\left[\mathfrak{r}_{B,N}(\h,x,y,z)\mathfrak{Z}^B_N(\h,x,y,z)+\mathfrak{R}_{B,N}(\h,x,y,z)\right]\phi(x-\h y,\xi-\eta)\psi(x-\h z,\xi-\zeta)+
$$
$$
+\pi^{-2d}\int_{\Xi}dY\hspace{-0.2cm}\int_{\Xi}dZ \ e^{-2i\sigma(Y,Z)}\left\{\sum_{0\leq\rho\leq N}\frac{(-i\h)^\rho}{\rho!}\left[\sum_{0\leq\lambda\leq N}\frac{\h^\lambda}{4\lambda!}\left(\frac{i}{2}\right)^\lambda\mathfrak{T}^B_\lambda(x)\right]^\rho\right\}\times
$$
$$
\times\left[\left(\frac{i}{2}\right)^{N+1}\hspace{-0.7cm}\sum_{|\alpha|=N+1}\left(\widetilde{\mathfrak{R}}_{\phi,N,\alpha}\big(\partial_\zeta^\alpha\psi\big)(x-\h z,\xi-\zeta)+(-1)^{N+1}\widetilde{\mathfrak{R}}_{\psi,N,\alpha}\big(\partial_\eta^\alpha\phi\big)(x-\h y,\xi-\eta)\right)-\right.
$$
$$
\left.-(-1)^{N+1}\left(\frac{i}{2}\right)^{2(N+1)}\hspace{-0.7cm}\sum_{|\alpha|=N+1}\sum_{|\beta|=N+1}\big(\partial_\eta^\beta\widetilde{\mathfrak{R}}_{\phi,N,\alpha}\big)(\h,x,y,z,\xi,\eta) \big(\partial_\zeta^\alpha\widetilde{\mathfrak{R}}_{\psi,N,\beta}\big)(\h,x,y,z,\xi,\zeta)\right]
$$

Let us discuss the remainders in the above expansion. First let us consider the factor $\mathfrak{Z}^B_N(\h,x,y,z)$. 
$$
\mathfrak{Z}^B_N(\h,x,y,z)=\sum_{1\leq\rho\leq N}\frac{(-i\h)^\rho}{\rho!}\sum_{0\leq\kappa\leq \rho-1}C^\kappa_\rho\mathfrak{r}_{B,N}(\h,x,y,z)^{\rho-\kappa}\left[\sum_{0\leq\lambda\leq N}\frac{\h^\lambda}{4\lambda!}\left(\frac{i}{2}\right)^\lambda\mathfrak{T}^B_\lambda(x)\right]^\kappa
$$
and taking into account the differential operators contained in $\mathfrak{T}^B_\lambda(x)$ (see (\ref{rest-T})) let us compute for $\rho\geq1$ and $0\leq\kappa\leq\rho-1$ one of the terms appearing in the above sum (with $0\leq\lambda_j\leq N$ for $1\leq j\leq\kappa$)
$$
\int_{\Xi}dY\hspace{-0.2cm}\int_{\Xi}dZ \ e^{-2i\sigma(Y,Z)}\mathfrak{r}_{B,N}(\h,x,y,z)^{\rho-\kappa}\mathfrak{T}^B_{\lambda_1}(x)\cdots\mathfrak{T}^B_{\lambda_\kappa}(x)\phi(x-\h y,\xi-\eta)\psi(x-\h z,\xi-\zeta)$$
$$
=\h^{(N+1)(\rho-\kappa)}\int_{\Xi}dY\hspace{-0.2cm}\int_{\Xi}dZ \ e^{-2i\sigma(Y,Z)}\left[\sum_{jk}y_jz_k\hspace{-0.4cm}\sum_{|\gamma|=N+1}\hspace{-0.2cm}(\gamma!)^{-1}F^B_{\gamma,j,k}(\h,x,y,z)\sum_{\delta\leq\gamma}\boldsymbol{T}^\gamma_\delta\ y^\delta z^{\gamma-\delta}\right]^{\rho-\kappa}\times
$$
$$
\times\mathfrak{T}^B_{\lambda_1}(x)\cdots\mathfrak{T}^B_{\lambda_\kappa}(x)\phi(x-\h y,\xi-\eta)\psi(x-\h z,\xi-\zeta)$$
$$
=\left(\frac{i}{2}\right)^{N+1}\h^{(N+1)(\rho-\kappa)}\sum_{|\alpha|=|\beta|=\rho-\kappa}\int_{\Xi}dY\hspace{-0.2cm}\int_{\Xi}dZ \ e^{-2i\sigma(Y,Z)}G^B_{\alpha\beta}(\h,x,y,z)\times
$$
\begin{equation}\label{rest-1}
\times\sum_{\delta\leq\gamma}(-1)^{N+1-|\delta|}\boldsymbol{T}^\gamma_\delta\partial_\zeta^\delta \partial_\eta^{\gamma-\delta}\mathfrak{T}^B_{\lambda_1}(x)\cdots\mathfrak{T}^B_{\lambda_\kappa}(x)\big(\partial_\xi^\beta\phi\big)(x-\h y,\xi-\eta)\big(\partial_\xi^\alpha\psi\big)(x-\h z,\xi-\zeta)
\end{equation}
where $G^B_{\alpha\beta}$ is a product of $\rho-\kappa$ functions of class $BC^\infty(\mathcal{X}^3)$ uniformly for $\h\in(0,\h_0]$, depending only on the derivatives of order $N+1$ of the magnetic field $B$. Moreover $\partial_\xi^\beta\phi\in S^{m_1-|\beta|}_1(\mathcal{X})$ and $\partial_\xi^\alpha\psi\in S^{m_2-|\alpha|}_1(\mathcal{X})$ uniformly in $\h\in(0,\h_0]$ so that the integral (\ref{rest-1}) defines an element in $S^{(N+1,m_1+m_2-2-(N+1))}_1(\mathcal{X})$ for any $\rho-\kappa\geq1$.

Let us consider the second contribution:
\begin{equation}\label{rest-2}
\int_{\Xi}dY\hspace{-0.2cm}\int_{\Xi}dZ \ e^{-2i\sigma(Y,Z)}\mathfrak{R}_{B,N}(\h,x,y,z)\phi(x-\h y,\xi-\eta)\psi(x-\h z,\xi-\zeta)\end{equation}
$$
=(-i\h)^{N+1}\left(\int_0^1du_1\ldots\int_0^{u_N}e^{-iu_{N+1}\h \theta^B_\h(x,y,z))}du_{N+1}\right)\times
$$
$$
\times\int_{\Xi}dY\hspace{-0.2cm}\int_{\Xi}dZ \ e^{-2i\sigma(Y,Z)}\left[\theta^B_\h(x,y,z))\right]^{N+1}\phi(x-\h y,\xi-\eta)\psi(x-\h z,\xi-\zeta)
$$
so that a similar procedure with the one used for (\ref{rest-1}) proves that this integral defines a function of class $S^{(N+1,m_1+m_2-2(N+1))}_1(\mathcal{X})$.

For any $\phi\in S^m(\mathcal{X})$ the rest $\widetilde{\mathfrak{R}}_{\phi,N,\alpha}$ is a function of class $S^{(N+1,m-(N+1))}_1(\mathcal{X})$, $\forall\alpha\in\mathbb{N}^d$ so that it is easy to notice that the last contribution to the rest:
$$
+\pi^{-2d}\int_{\Xi}dY\hspace{-0.2cm}\int_{\Xi}dZ \ e^{-2i\sigma(Y,Z)}\left\{\sum_{0\leq\rho\leq N}\frac{(-i\h)^\rho}{\rho!}\left[\sum_{0\leq\lambda\leq N}\frac{\h^\lambda}{4\lambda!}\left(\frac{i}{2}\right)^\lambda\mathfrak{T}^B_\lambda(x)\right]^\rho\right\}\times
$$
$$
\times\left[\left(\frac{i}{2}\right)^{N+1}\hspace{-0.7cm}\sum_{|\alpha|=N+1}\left(\widetilde{\mathfrak{R}}_{\phi,N,\alpha}\big(\partial_\zeta^\alpha\psi\big)(x-\h z,\xi-\zeta)+(-1)^{N+1}\widetilde{\mathfrak{R}}_{\psi,N,\alpha}\big(\partial_\eta^\alpha\phi\big)(x-\h y,\xi-\eta)\right)-\right.
$$
$$
\left.-(-1)^{N+1}\left(\frac{i}{2}\right)^{2(N+1)}\hspace{-0.7cm}\sum_{|\alpha|=N+1}\sum_{|\beta|=N+1}\big(\partial_\eta^\beta\widetilde{\mathfrak{R}}_{\phi,N,\alpha}\big)(\h,x,y,z,\xi,\eta) \big(\partial_\zeta^\alpha\widetilde{\mathfrak{R}}_{\psi,N,\beta}\big)(\h,x,y,z,\xi,\zeta)\right]
$$
defines an element of $S^{(N+1,m_1+m_2-2(N+1))}_1(\mathcal{X})$.

Let us now concentrate on the main terms in the expansion (\ref{main-dev}):
$$
\pi^{-2d}\int_{\Xi}dY\hspace{-0.2cm}\int_{\Xi}dZ \ e^{-2i\sigma(Y,Z)}\left\{\sum_{0\leq\rho\leq N}\frac{(-i\h)^\rho}{\rho!}\left[\sum_{0\leq\lambda\leq N}\frac{\h^\lambda}{4\lambda!}\left(\frac{i}{2}\right)^\lambda\mathfrak{T}^B_\lambda(x)\right]^\rho\right\}\times
$$
$$
\times\left[\sum_{0\leq\nu\leq N}\frac{(-\h)^\nu}{\nu!}\sum_{|\alpha|=\nu}\frac{\nu!}{\alpha!}\left((i/2)\partial_\zeta\right)^\alpha\big(\partial_x^\alpha \phi\big)(x,\xi-\eta)\right]\left[\sum_{0\leq\mu\leq N}\frac{(-\h)^\mu}{\mu!}\sum_{|\beta|=\mu}\frac{\mu!}{\beta!}\left((-i/2)\partial_\eta\right)^\beta\big(\partial_x^\beta \psi\big)(x,\xi-\zeta)\right]$$
$$
=\pi^{-2d}\int_{\Xi}dY\hspace{-0.2cm}\int_{\Xi}dZ \ e^{-2i\sigma(Y,Z)}\left\{\sum_{0\leq\rho\leq N}\frac{(-i\h)^\rho}{4^\rho\rho!}\right.\underset{\begin{array}{c}\scriptstyle{\{\lambda_1,\ldots,\lambda_\rho\}}\\\scriptstyle{0\leq \lambda_j\leq \rho}\\\scriptstyle{1\leq j\leq \rho}\end{array}}{\sum}\left.\left(\frac{i\h}{2}\right)^{\lambda_1+\cdots+\lambda_\rho}\frac{1}{\lambda_1!\cdots\lambda_\rho!}\mathfrak{T}^B_{\lambda_1}(x)\cdots\mathfrak{T}^B_{\lambda_\rho}(x) \right\}\times
$$
$$
\times\underset{\begin{array}{c}\scriptstyle{0\leq\nu\leq N}\\\scriptstyle{0\leq\mu\leq N}\end{array}}{\sum}(-1)^\mu\left(\frac{i\h}{2}\right)^{\nu+\mu}
\underset{\begin{array}{c}\scriptstyle{|\alpha|=\nu}\\\scriptstyle{|\beta|=\mu}\end{array}}{\sum}\left[\frac{1}{\alpha!\beta!}\big(\partial_\xi^\beta\partial_x^\alpha\phi\big)(x,\xi-\eta)\big(\partial_\xi^\alpha\partial_x^\beta\psi\big)(x,\xi-\zeta) \right]$$
\begin{equation}\label{c}
 =\sum_{0\leq k\leq N}\h^k\underset{\begin{array}{c}\scriptstyle{l_1+l_2+l=k}\\\scriptstyle{0\leq l_1 \leq k}\\\scriptstyle{0\leq l_2 \leq k}\\\scriptstyle{0\leq l \leq k}\end{array}}{\sum}\underset{\begin{array}{c}\scriptstyle{\rho+\lambda_1+\ldots+\lambda_\rho=l}\\\scriptstyle{0\leq \rho \leq l}\\\scriptstyle{0\leq \lambda_j \leq l}\\\scriptstyle{1\leq j \leq \rho}\end{array}}{\sum}\frac{(-1)^{\rho+l_2}i^\rho}{4^\rho\rho!}
\underset{\begin{array}{c}\scriptstyle{|\alpha|=l_1}\\\scriptstyle{|\beta|=l_2}\end{array}}{\sum}\left(\frac{i}{2}\right)^{l_1+l_2+\lambda_1+\ldots+\lambda_\rho}\frac{1}{\alpha!\beta!}\times
\end{equation}
$$
\times\underset{\begin{array}{c}\scriptstyle{|\gamma_1|=\lambda_1}\\\scriptstyle{\delta_1\leq\gamma_1}\end{array}}{\sum}\ldots\underset{\begin{array}{c}\scriptstyle{|\gamma_\rho|=\lambda_\rho}\\\scriptstyle{\delta_\rho\leq\gamma_\rho}\end{array}}{\sum}(-1)^{|\delta_1|+\ldots+|\delta_\rho|}T^{\gamma_1}_{\delta_1}\cdots T^{\gamma_\rho}_{\delta_\rho}\big(\partial^{\gamma_1}B_{j_1 k_1}\big)(x)\cdots\big(\partial^{\gamma_\rho}B_{j_\rho k_\rho}\big)(x)\times
$$
$$
\times\left(\partial_{\xi_{k_1}}\cdots\partial_{\xi_{k_\rho}} \partial_\xi^{\beta+\gamma_1-\delta_1+\ldots+\gamma_\rho-\delta_\rho}\partial_x^\alpha\phi\right)(x,\xi)\left(\partial_{\xi_{j_1}}\cdots\partial_{\xi_{j_\rho}} \partial_\xi^{\alpha+\delta_1+\ldots+\delta_\rho}\partial_x^\beta\psi\right)(x,\xi)\ +\ R_N(\h;\phi,\psi)(x,\xi),
$$
where $R_N\in S^{(N+1,m_1+m_2-(N+1))}_1(\mathcal{X})$.

Using these expansions one gets the following statement.
\begin{proposition}\label{h-B-dev}~\\
If the magnetic field $B$ has components of class $BC^\infty(\mathcal{X})$ and $\phi\in S^{m_1}_1(\mathcal{X})$, $\psi\in S^{m_2}_1(\mathcal{X})$ then there exists a sequence $\{\mathfrak{c}^B_k(\phi,\psi)\}_{k\in\mathbb{N}}$ such that $\mathfrak{c}^B_k(\phi,\psi)\in S^{m_1+m_2-k}_1(\mathcal{X})$ and, for any $N\geq1$ in $\mathbb{N}$, we have:
$$
\h^{-N}\left(\phi\sharp^B_{\hbar}\psi-\sum_{k=0}^{N-1}\h^k\mathfrak{c}^B_k(\phi,\psi)\right)\in S^{m_1+m_2-N}_1(\mathcal{X})
$$
uniformly for $\h\in\I$. For $k=0$, we have $\mathfrak{c}^B_0(\phi,\psi)=\phi\cdot\psi$. Moreover, for any $k\in\mathbb{N}$, the function $\mathfrak{c}^B_k(\phi,\psi;X)$ only depends on values of the magnetic field $B$ and its derivatives of order $\leq j-1$ evaluated at $x$.
\end{proposition}
We shall use the notation
\begin{equation}\label{dev-gamma}
\gamma^B_\h(\phi,\psi):=\phi\sharp^B_\h\psi-\phi\cdot\psi=\sum_{k=1}^{N-1}\h^k\mathfrak{c}^B_k(\phi,\psi)+\h^N\mathfrak{t}^B_N(\phi,\psi),
\end{equation}
with $\mathfrak{t}^B_N(\phi,\psi)\in S^{(0,m_1+m_2-N)}_1(\mathcal{X})$.

Starting from the above result we shall work in the sequel with functions of class $S^{(s,m)}_1(\mathcal{X})$, thus admiting asymptotic expansions in $\h\in\I$ of the form given in the above proposition.

\section{The magnetic Schr\"{o}dinger operator}\label{B-V-Lapl}

\subsection{Preliminaries}

Let us notice that the magnetic Schr\"odinger operator defined in (\ref{magn-Lapl}) satisfies
\begin{equation}\label{Op-A-h}
P^{A}(\h)=\mathfrak{Op}^A_\h(\xi^2+V),.
\end{equation}
We shall suppose that the magnetic
field has components of class $BC^\infty(\mathcal{X})$, that the
vector potential has been chosen of class $C^\infty_{\text{\sf
    pol}}(\mathcal{X})$ and that $V$ is a real
$BC^\infty(\mathcal{X})$ function. Hence  $F:=\xi^2+V\in
S^{2}_1(\mathcal{X})$ and all the results in Section \ref{B-h-calc}
can be applied in order to conclude that (taking also into account
that $V$ is a bounded self-adjoint perturbation of $P^A_0(\h):=\mathfrak{Op}^A_\h(\xi^2)$ and the
Neumann series expansion of the resolvent)~:
\begin{proposition}\label{Prop-P-A-h}~
\begin{enumerate}
 \item $P^{A}(\h)$ defined in (\ref{Op-A-h}) is essentially
   self-adjoint on $\mathcal{S}(\mathcal{X})$ and its self-adjoint
   extension, denoted $H$, has the domain $\mathcal{H}^A_2(\mathcal{X})$.
\item The resolvent $(H-\mathfrak{z})^{-1}$ is well defined for
  $\mathfrak{z}\in\mathbb{C}\setminus\{\mathfrak{x}\in\mathbb{R}\mid\mathfrak{x}\geq\mathfrak{a}\}$
  for some $\mathfrak{a}\in \mathbb R$ and has the form $(H-\mathfrak{z})^{-1}=\mathfrak{Op}^A_\h(r^B_\mathfrak{z})$ with $r^B_\mathfrak{z}\in S^{(0,-2)}_1(\mathcal{X})$.
\item There exists $\mathfrak{a}_0\geq\mathfrak{a}$ depending on $B$ and $V$ such that $\sigma_{\text{\sf ess}}(H)\subset[\mathfrak{a}_0,\infty)$.
\end{enumerate}
\end{proposition}

\subsection{The resolvent}

We shall concentrate now on the asymptotic expansion of the symbol $r^B_\mathfrak{z}\in S^{(0,-2)}_1(\mathcal{X})$ with respect to $\h\in\I$. In fact, as the case witout magnetic field is well-known, we shall only be interested in the 'magnetic' contribution to the terms of the $\h$-asymptotic expansion, specifically in putting them in a manifestly gauge invariant form. For this purpose we shall use our parametrix type construction of \cite{MPR} in order to express $r^B_\mathfrak{z}$ in terms of $(F-\mathfrak{z})^{-1}$. In fact we know that
$
(F-\mathfrak{z})\sharp^B_\h r^B_\mathfrak{z}=1.
$
In order to shorten our notations we shall denote by $p_{_\mathfrak{z}}:=F-\mathfrak{z}$. Let us compute
$$
\p\sharp^B_\h\p^{-1}=1\,+\,\gamma^B_\h\big(\p,\p^{-1}\big)=1+\sum_{k=1}^{N-1}\h^k\mathfrak{c}^B_k(\p,\p^{-1})+\h^{N}\mathfrak{t}^B_{N}(\p,\p^{-1})
$$
with $\mathfrak{c}^B_k(\p,\p^{-1})\in S^{-j}_1(\mathcal{X})$ and $\mathfrak{t}^B_{N}(\p,\p^{-1})\in S^{(0,-N)}_1(\mathcal{X})$. Thus
$$
r^B_\mathfrak{z}=\p^{-1}-r^B_\mathfrak{z}\sharp^B_\h\gamma^B_\h\big(\p,\p^{-1}\big)$$
\begin{equation}\label{rB-dev}
=\sum_{0\leq j\leq M-1}(-1)^j\p^{-1}\sharp^B_\h\gamma^B_\h\big(\p,\p^{-1}\big)^{\sharp^B_\h j}+(-1)^Mr^B_\mathfrak{z}\sharp^B_\h\gamma^B_\h\big(\p,\p^{-1}\big)^{\sharp^B_\h M}\end{equation}
$$
=\sum_{0\leq j\leq M-1}(-1)^j\p^{-1}\sharp^B_\h\left[\sum_{k=1}^{N_1-1}\h^k\mathfrak{c}^B_k(\p,\p^{-1})+\h^{N_1}\mathfrak{t}^B_{N_1}(\p,\p^{-1})\right]^{\sharp^B_\h j}+
$$
$$
+(-1)^Mr^B_\mathfrak{z}\sharp^B_\h\left[\sum_{k=1}^{N_2-1}\h^k\mathfrak{c}^B_k(\p,\p^{-1})+\h^{N_2}\mathfrak{t}^B_{N_2}(\p,\p^{-1})\right]^{\sharp^B_\h M}.
$$

Let us first study the corrections $\mathfrak{c}^B_k(\p,\p^{-1})$. 
We use the general formulae (\ref{dev-gamma}), (\ref{c})  and obtain:
$$
\mathfrak{c}^B_k(\p,\p^{-1})=(2\pi)^{-2d}\underset{\begin{array}{c}\scriptstyle{l_1+l_2+l=k}\\\scriptstyle{0\leq l_1 \leq k}\\\scriptstyle{0\leq l_2 \leq k}\\\scriptstyle{0\leq l \leq k}\end{array}}{\sum}\underset{\begin{array}{c}\scriptstyle{\rho+\lambda_1+\ldots+\lambda_\rho=l}\\\scriptstyle{0\leq \rho \leq l}\\\scriptstyle{0\leq \lambda_j \leq l}\\\scriptstyle{1\leq j \leq \rho}\end{array}}{\sum}\frac{(-1)^{\rho+l_2}i^\rho}{4^\rho\rho!}
\underset{\begin{array}{c}\scriptstyle{|\alpha|=l_1}\\\scriptstyle{|\beta|=l_2}\end{array}}{\sum}\left(\frac{i}{2}\right)^{l_1+l_2+\lambda_1+\ldots+\lambda_\rho}\frac{1}{\alpha!\beta!}\times
$$
$$
\times\underset{\begin{array}{c}\scriptstyle{|\gamma_1|=\lambda_1}\\\scriptstyle{\delta_1\leq\gamma_1}\end{array}}{\sum}\ldots\underset{\begin{array}{c}\scriptstyle{|\gamma_\rho|=\lambda_\rho}\\\scriptstyle{\delta_\rho\leq\gamma_\rho}\end{array}}{\sum}(-1)^{|\delta_1|+\ldots+|\delta_\rho|}T^{\gamma_1}_{\delta_1}\cdots T^{\gamma_\rho}_{\delta_\rho}\big(\partial^{\gamma_1}B_{j_1 k_1}\big)(x)\cdots\big(\partial^{\gamma_\rho}B_{j_\rho k_\rho}\big)(x)\times
$$
$$
\times\left(\partial_{\xi_{k_1}}\cdots\partial_{\xi_{k_\rho}} \partial_\xi^{\beta+\gamma_1-\delta_1+\ldots+\gamma_\rho-\delta_\rho}\partial_x^\alpha\p\right)(x,\xi)\left(\partial_{\xi_{j_1}}\cdots\partial_{\xi_{j_\rho}} \partial_\xi^{\alpha+\delta_1+\ldots+\delta_\rho}\partial_x^\beta(\p^{-1})\right)(x,\xi).
$$
An important observation is that all the derivatives of the symbols $\p$ and $\p^{-1}$ have a very special dependence on $\mathfrak{z}$. More precisely we have:
\begin{itemize}
 \item $
\big(\partial_{x}^\alpha p_{\mathfrak{z}}\big)(x,\xi)=\big(\partial_{x}^\alpha F\big)(x,\xi)=\big(\partial_{x}^\alpha V\big)(x)\,;
$
\item $
\big(\partial_{\xi_j}p_{\mathfrak{z}}\big)(x,\xi)=\big(\partial_{\xi_j}F\big)(x,\xi)=2\xi_j;\quad
\big(\partial_{\xi_j}\partial_{\xi_k}p_{\mathfrak{z}}\big)(x,\xi)=2\delta_{jk};\quad
\big(\partial_{\xi}^\alpha p_{\mathfrak{z}}\big)(x,\xi)=0,\quad\forall|\alpha|\geq3\,
$;
\item $
\big(\partial_{x}^\alpha\partial_{\xi}^\beta
p_{\mathfrak{z}}\big)(x,\xi)=0,\quad\text{for
}|\alpha|\geq1,|\beta|\geq 1\,.
$
\end{itemize}
\begin{lemma}\label{p-inv-deriv}
For any multiindices $\alpha$ and $\beta$ we have that:
$$
\big(\partial^\alpha_x\partial^\beta_\xi\p\big)(x,\xi)=\sum_{0\leq k\leq|\alpha|+|\beta|}\mathfrak{q}_k(x,\xi)\p^{-1-k}(x,\xi)
$$
where $\mathfrak{q}_k(x,\xi)$ are polynomials of degree at most $k$ in $\xi$ with coefficients functions of $x$ depending only on the first $|\alpha|$ derivatives of $V(x)$.
\end{lemma}
\begin{proof}~\\
In fact we have:\\
$
\big(\partial_{x_j}p^{-1}_{\mathfrak{z}}\big)(x,\xi)=-p^{-2}_{\mathfrak{z}}(x,\xi)\big(\partial_{x_j}F\big)(x,\xi)=-p^{-2}_{\mathfrak{z}}(x,\xi)\big(\partial_{x_j}V\big)(x);
$\\
 $
\big(\partial_{\xi_j}p^{-1}_{\mathfrak{z}}\big)(x,\xi)=-p^{-2}_{\mathfrak{z}}(x,\xi)\big(\partial_{\xi_j}F\big)(x,\xi)=-2\xi_jp^{-2}_{\mathfrak{z}}(x,\xi).
$\\
Thus the statement of the Lemma is true for $|\alpha|+|\beta|=1$ and we shall proceed by induction on $|\alpha|+|\beta|$. Suppose the statement has been proved for $|\alpha|+|\beta|=N$ and let us compute the next derivatives.
$$
\big(\partial_{x_j}\partial^\alpha_x\partial^\beta_\xi\p\big)(x,\xi)=\partial_{x_j}\left[\sum_{0\leq k\leq N}\mathfrak{q}_k\p^{-1-k}\right](x,\xi)$$
$$
=\sum_{0\leq k\leq N}\big(\partial_{x_j}\mathfrak{q}_k\big)(x,\xi)\p^{-1-k}(x,\xi)-(1+k)\sum_{0\leq k\leq N}\mathfrak{q}_k(x,\xi)\big(\partial_{x_j}p_{\mathfrak{z}}\big)(x,\xi)\p^{-1-(k+1)}(x,\xi)$$
$$
=\sum_{0\leq k\leq N}\left[\big(\partial_{x_j}\mathfrak{q}_k\big)(x,\xi)-k \mathfrak{q}_{k-1}(x,\xi)\big(\partial_{x_j}V\big)(x)\right]\p^{-1-k}(x,\xi)\ -\ (1+N)\mathfrak{q}_{N}(x,\xi)\big(\partial_{x_j}V\big)(x)\p^{-1-(N+1)}(x,\xi);
$$
$$
\big(\partial_{\xi_j}\partial^\alpha_x\partial^\beta_\xi\p\big)(x,\xi)=\partial_{\xi_j}\left[\sum_{0\leq k\leq N}\mathfrak{q}_k\p^{-1-k}\right](x,\xi)$$
$$
=\sum_{0\leq k\leq N}\big(\partial_{\xi_j}\mathfrak{q}_k\big)(x,\xi)\p^{-1-k}(x,\xi)-(1+k)\sum_{0\leq k\leq N}\mathfrak{q}_k(x,\xi)\big(\partial_{\xi_j}p_{\mathfrak{z}}\big)(x,\xi)\p^{-1-(k+1)}(x,\xi)$$
$$
=\sum_{0\leq k\leq N}\left[\big(\partial_{\xi_j}\mathfrak{q}_k\big)(x,\xi)-2k\xi_j \mathfrak{q}_{k-1}(x,\xi)\right]\p^{-1-k}(x,\xi)\ -\ 2(1+N)\xi_j\mathfrak{q}_{N}(x,\xi)\p^{-1-(N+1)}(x,\xi).
$$
Thus the formula is valid also for $|\alpha|+|\beta|=N+1$ with coefficients having obviously the same structure.
\end{proof}
\begin{remark}
Some simple computation proves that:
$$
\mathfrak{c}^B_1(\p,\p^{-1})=0,
$$
$$
\mathfrak{c}^B_2(\p,\p^{-1})=\frac{1}{2}p^{-2}_{\mathfrak{z}}(x,\xi)\big(\Delta V\big)(x)-\frac{1}{2}p^{-3}_{\mathfrak{z}}(x,\xi)\left|\big(\nabla V\big)(x)\right|^2-
$$
$$
-2p^{-3}_{\mathfrak{z}}(x,\xi)\sum_{lm}\left(1-\frac{1}{2}\delta_{lm}\right)\big(\partial_{x_l}\partial_{x_m}V\big)(x)\xi_l\xi_m+
$$
$$
+\frac{1}{2}p^{-2}_{\mathfrak{z}}(x,\xi)\left|B(x)\right|^2 
-2p^{-3}_{\mathfrak{z}}(x,\xi)\sum_{jkm}B_{jk}(x)B_{jm}(x)\xi_k\xi_m+\frac{2}{3}p^{-2}_{\mathfrak{z}}(x,\xi)\sum_{jk}\left(\partial_jB_{jk}\right)(x)\xi_k-
$$
$$
-2p^{-3}_{\mathfrak{z}}(x,\xi)\sum_{jk}B_{jk}(x)\xi_k\big(\partial_{x_j}V\big)(x).
$$
\end{remark}
Developing successively each $\sharp^B_\h$-product in (\ref{rB-dev}) 
and using Proposition \ref{h-B-dev}, one gets the following statement.
\begin{proposition}\label{h-dev-resolvent}~\\
The 'magnetic' symbol $r^B_\mathfrak{z}$ admits for any $N\in\mathbb{N}$ an asymptotic expansion in $\h$ of the form
$$
r^B_\mathfrak{z}(X)=\p^{-1}+\sum_{1\leq j\leq N-1}\h^j\mathfrak{r}^B_j(\mathfrak{z};X)+\h^{N}\widetilde{\mathfrak{r}^B_N}(\mathfrak{z};\h,X)
$$
where the terms $\mathfrak{r}^B_j(\mathfrak{z};X)$ only depend on the magnetic field $B$ and its derivatives up to order $j-1$ evaluated at $X$, and the rest $\widetilde{\mathfrak{r}^B_N}(\mathfrak{z};\h,X)$ only depends on the magnetic field $B$ (in a non-local way). Moreover $\mathfrak{r}^B_j(\mathfrak{z})\in S^{-j}_1(\mathcal{X})$ and $\widetilde{\mathfrak{r}^B_N}(\mathfrak{z})\in S^{(0,-N)}_1(\mathcal{X})$.
\end{proposition}
Considering (\ref{rB-dev}) we obtain the following expansion of $r^B_\mathfrak{z}$ in powers of $\h$
\begin{equation}\label{remainder}
r^B_\mathfrak{z}\sim\p^{-1}+\sum_{1\leq j}\h^j\sum_{1\leq k \leq j}(-1)^k\underset{\begin{array}{c}\scriptstyle{\lambda_1+\ldots+\lambda_k=j}\\\scriptstyle{1\leq \lambda_l \leq j}\\\scriptstyle{1\leq l \leq k}\end{array}}{\sum}\p^{-1}\sharp^B_\h\mathfrak{c}^B_{\lambda_1}(\p,\p^{-1})\sharp^B_\h\ldots\sharp^B_\h\mathfrak{c}^B_{\lambda_k}(\p,\p^{-1})
\end{equation}
 and developing further each $\h$-dependent $\sharp^B_\h$-product, one has
$$
r^B_\mathfrak{z}\sim\p^{-1}+\sum_{1\leq j}\sum_{1\leq k \leq j}(-1)^k\underset{\begin{array}{c}\scriptstyle{\lambda_1+\ldots+\lambda_k=j}\\\scriptstyle{1\leq \lambda_l \leq j}\\\scriptstyle{1\leq l \leq k}\end{array}}{\sum}\underset{\begin{array}{c}\scriptstyle{0\leq \mu_l}\\\scriptstyle{1\leq l \leq k}\end{array}}{\sum}\h^{j+\mu_1+\ldots+\mu_k}\mathfrak{C}^B_{\mu_k,\ldots\mu_1}\big(\p^{-1},\mathfrak{c}^B_{\lambda_1}(\p,\p^{-1}),\ldots,\mathfrak{c}^B_{\lambda_k}(\p,\p^{-1})\big)$$
$$
=\p^{-1}+\sum_{1\leq n}\h^n\sum_{1\leq j\leq n}\sum_{1\leq k \leq j}(-1)^k\underset{\begin{array}{c}\scriptstyle{\lambda_1+\ldots+\lambda_k=j}\\\scriptstyle{1\leq \lambda_l \leq j}\\\scriptstyle{1\leq l \leq k}\end{array}}{\sum}\underset{\begin{array}{c}\scriptstyle{j+\mu_1+\ldots+\mu_k=n}\\\scriptstyle{0\leq \mu_l}\\\scriptstyle{1\leq l \leq k}\end{array}}{\sum}\mathfrak{C}^B_{\mu_k,\ldots\mu_1}\big(\p^{-1},\mathfrak{c}^B_{\lambda_1}(\p,\p^{-1}),\ldots,\mathfrak{c}^B_{\lambda_k}(\p,\p^{-1})\big)
$$
where
$$
\mathfrak{C}^B_{\mu_k,\ldots\mu_1}\big(f,g_1,\ldots,g_k\big):=\mathfrak{c}^B_{\mu_k}(\mathfrak{c}^B_{\mu_{k-1}}(\ldots\mathfrak{c}^B_{\mu_1}(f,g_1),g_2),\ldots,g_k)
$$

Putting together all these formulae we conclude that:
\begin{proposition}\label{str-r-j}~\\
Each term $\mathfrak{r}^B_j(\mathfrak{z})$ for $j\geq1$ is a finite sum of the form
$$
\mathfrak{r}^B_j(\mathfrak{z})=\sum_{0\leq p\leq j}\mathfrak{f}^B_p(x,\xi)\p^{-2-p}(x,\xi)
$$
where $\mathfrak{f}^B_p(x,\xi)$ are polynomials
 in $\xi$ of degree at most $p$ whose coefficients
 are  $C^\infty$ functions of $x$ depending only
on a finite number of partial derivatives of $V$ and $B$ at the given point $x$.
\end{proposition}

\begin{remark}
Some tedious computation gives:\\
$
\begin{array}{lcl}\mathfrak{r}^B_0(\mathfrak{z})=\p^{-1}\end{array}$;\\
$\begin{array}{lcl}\mathfrak{r}^B_1(\mathfrak{z})=-\p^{-1}\mathfrak{c}^B_1(\p,\p^{-1})=0\end{array}$;\\
$
\begin{array}{lcl}
\mathfrak{r}^B_2(\mathfrak{z})&=&-\p^{-1}\mathfrak{c}^B_2(\p,\p^{-1})+\p^{-1}\mathfrak{c}^B_1(\p,\p^{-1})\mathfrak{c}^B_1(\p,\p^{-1})-\mathfrak{c}^B_1\big(\p^{-1},\mathfrak{c}^B_1(\p,\p^{-1})\big)\\
&=&-\p^{-1}\mathfrak{c}^B_2(\p,\p^{-1})\\
&=&-\frac{1}{2}p^{-3}_{\mathfrak{z}}(x,\xi)\big(\Delta V\big)(x)+\frac{1}{2}p^{-4}_{\mathfrak{z}}(x,\xi)\left|\big(\nabla V\big)(x)\right|^2\\
&&+2p^{-4}_{\mathfrak{z}}(x,\xi)\sum_{lm}\left(1-\frac{1}{2}\delta_{lm}\right)\big(\partial_{x_l}\partial_{x_m}V\big)(x)\xi_l\xi_m\\
&&-\frac{1}{2}p^{-3}_{\mathfrak{z}}(x,\xi)\left|B(x)\right|^2 
+2p^{-4}_{\mathfrak{z}}(x,\xi)\sum_{jkm}B_{jk}(x)B_{jm}(x)\xi_k\xi_m-\frac{2}{3}p^{-3}_{\mathfrak{z}}(x,\xi)\sum_{jk}\left(\partial_jB_{jk}\right)(x)\xi_k\\
&&+2p^{-4}_{\mathfrak{z}}(x,\xi)\sum_{jk}B_{jk}(x)\xi_k\big(\partial_{x_j}V\big)(x);
\end{array}
$\\
$
\begin{array}{lcl}
\mathfrak{r}^B_3(\mathfrak{z})&=&-\p^{-1}\mathfrak{c}^B_3(\p,\p^{-1})+2\p^{-1}\mathfrak{c}^B_1(\p,\p^{-1})\mathfrak{c}^B_2(\p,\p^{-1})-\p^{-1}\big[\mathfrak{c}^B_1(\p,\p^{-1})\big]^3+\p^{-1}\mathfrak{c}^B_1\big(\mathfrak{c}^B_1(\p,\p^{-1}),\mathfrak{c}^B_1(\p,\p^{-1})\big)\\
&&-\mathfrak{c}^B_1\big(\p^{-1},\mathfrak{c}^B_2(\p,\p^{-1})\big)-\mathfrak{c}^B_2\big(\p^{-1},\mathfrak{c}^B_1(\p,\p^{-1})\big)+\mathfrak{c}^B_1\big(\p^{-1},\big[\mathfrak{c}^B_1(\p,\p^{-1})\big]^2\big)\\
&=&-\p^{-1}\mathfrak{c}^B_3(\p,\p^{-1})-\mathfrak{c}^B_1\big(\p^{-1},\mathfrak{c}^B_2(\p,\p^{-1})\big);
\end{array}
$
$
\begin{array}{lcl}
\mathfrak{r}^B_4(\mathfrak{z})&=&-\p^{-1}\mathfrak{c}^B_4(\p,\p^{-1})+2\p^{-1}\mathfrak{c}^B_1(\p,\p^{-1})\mathfrak{c}^B_3(\p,\p^{-1})+\p^{-1}\big[\mathfrak{c}^B_2(\p,\p^{-1})\big]^2-3\p^{-1}\big[\mathfrak{c}^B_1(\p,\p^{-1})\big]^2\mathfrak{c}^B_2(\p,\p^{-1})\\
&&+\p^{-1}\big[\mathfrak{c}^B_1(\p,\p^{-1})\big]^4-\mathfrak{c}^B_3\big(\p^{-1},\mathfrak{c}^B_1(\p,\p^{-1})\big)-\mathfrak{c}^B_2\big(\p^{-1},\mathfrak{c}^B_2(\p,\p^{-1})\big)\\
&&+2\mathfrak{c}^B_2\big(\p^{-1},\big[\mathfrak{c}^B_1(\p,\p^{-1})\big]^2\big)-\mathfrak{c}^B_1\big(\p^{-1},\mathfrak{c}^B_3(\p,\p^{-1})\big)\\
&&+2\mathfrak{c}^B_1\big(\p^{-1},\mathfrak{c}^B_1(\p,\p^{-1})\mathfrak{c}^B_2(\p,\p^{-1})\big)-\mathfrak{c}^B_1\big(\p^{-1},\big[\mathfrak{c}^B_1(\p,\p^{-1})\big]^3\big)+\mathfrak{c}^B_1\left(\p^{-1},\mathfrak{c}^B_1\big(\mathfrak{c}^B_1(\p,\p^{-1}),\mathfrak{c}^B_1(\p,\p^{-1})\big)\right)\\
&=&-\p^{-1}\mathfrak{c}^B_4(\p,\p^{-1})+\p^{-1}\big[\mathfrak{c}^B_2(\p,\p^{-1})\big]^2-\mathfrak{c}^B_2\big(\p^{-1},\mathfrak{c}^B_2(\p,\p^{-1})\big)-\mathfrak{c}^B_1\big(\p^{-1},\mathfrak{c}^B_3(\p,\p^{-1})\big).
\end{array}
$
\end{remark}

\subsection{The functional calculus with the magnetic Schr\"odinger operator}

Using the results recalled in Section \ref{B-h-calc} we conclude that for any function $g\in C^\infty_0(\mathbb{R})$, we can define by the functional calculus
 for  self-adjoint operators a bounded operator $g(H)$
 that can be computed using Formula  (\ref{DHS}). Then, using the fact that $(H-\mathfrak{z})^{-1}=\mathfrak{Op}^A_\h(r^B_\mathfrak{z})\,$, we conclude that we can compute directly the symbol of $g(H)=:\mathfrak{Op}^A_\h(\widetilde{g(F)}^B_\h)$,
\begin{equation}\label{B-f-calc}
\widetilde{g(F)}^B_\h(X) = -\pi^{-1} \lim_{\epsilon\ar 0^+} \int \!\!\int_{|\mu|\geq \epsilon}
\frac{\pa {\tilde g}}{\pa {\bar z}}(\lambda,\mu)\; r^B_{\lambda+i\mu}(X) d\lambda\,d\mu,\quad\forall X\in\Xi,
\end{equation}
defining a function of class $S^{(0,-2)}_1(\mathcal{X})$. Using our previous Proposition \ref{h-dev-resolvent} we obtain:
\begin{proposition}\label{h-dev-fcalc}~\\
The 'magnetic' symbol $\widetilde{g(F)}^B_\h$ admits for any $K\in\mathbb{N}$ an asymptotic expansion in $\h$ of the form
$$
\widetilde{g(F)}^B_\h(X)=\sum_{0\leq j\leq K-1}\h^j g^B_j[F](X)+\h^K\widetilde{g^B_K}[F](\h,X)
$$
where the terms $g^B_j[F](X)$ only depend on the magnetic field $B$ and its derivatives up to order $j-1$ evaluated at $X$, and the rest $\widetilde{g^B_K}[F](\h,X)$ only depends on the magnetic field $B$ (in a non-local way). Moreover $g^B_j[F]\in S^{-j}_1(\mathcal{X})$ and $\widetilde{g^B_K}[F]\in S^{(0,-K)}_1(\mathcal{X})$.
\end{proposition}

Let us consider a function $g\in C^\infty_0(\mathbb{R})$ such that its support $\Sigma_g$ is contained in $]-\infty,\mathfrak{a}_0[$. As the spectrum of $H$ is discrete in this region, and $\Sigma_g$ is a compact subset of $]-\infty,\mathfrak{a}_0[$, we deduce that $g(H)$ is finite rank and thus trace-class. Moreover, $g(H)=\mathfrak{Op}^A(\widetilde{g(F)}^B_\h)$ is an integral operator having the integral kernel (see \cite{MP})
$$
K^A[g(H)](x,y):=\left(e^{-\frac{i}{\h}\int_{[x,y]}A}\right)\mathfrak{F}^-\left[\widetilde{g(F)}^B_\h\right]\left(\frac{x+y}{2},x-y\right)
$$
with $\mathfrak{F}^-$ the inverse Fourier transform in the second variable (as defined on distributions on $\Xi=\mathcal{X}\times\mathcal{X}^\prime$). In order to study the regularity properties of this integral kernel we shall use our formula (\ref{B-f-calc}) and write that
$$
r^B_{\lambda+i\mu}(X)=(F-(\lambda+i\mu))^{-1}(X)+\mathfrak{X}^B_{\lambda+i\mu}(X)
$$
where
$$
-\pi^{-1} \lim_{\epsilon\ar 0^+} \int \!\!\int_{|\mu|\geq \epsilon}
\frac{\pa {\tilde g}}{\pa {\bar z}}(\lambda,\mu)\; (F-(\lambda+i\mu))^{-1} d\lambda\,d\mu=(g\circ F)\in C^\infty_0(\Xi)
$$
and
$$
\mathfrak{X}^B_{\lambda+i\mu}=r^B_{\lambda+i\mu}-(F-(\lambda+i\mu))^{-1}=r^B_{\lambda+i\mu}\sharp^B_\h\left(1-(F-(\lambda+i\mu))\sharp^B_\h(F-(\lambda+i\mu))^{-1}\right)\in S^{-2-2}_1(\mathcal{X})
$$
due to our Proposition \ref{P3} applied to $1-(F-(\lambda+i\mu))\sharp^B_\h(F-(\lambda+i\mu))^{-1}$. Let us remark that in 2 or 3 dimensions, $\mathfrak{F}^-S^{-4}_1(\mathcal{X})$ is contained in the space of jointly continuous functions on $\mathcal{X}\times\mathcal{X}$ (by the Riemann-Lebesgue Lemma). Let us recall that for trace-class operators with continuous integral kernels we have the following property.
\begin{proposition}~\\
Suppose $T\in\mathbb{B}_1\big(L^2(\mathcal{X})\big)$ has an integral kernel $K[T]\in C(\mathcal{X}\times\mathcal{X})$. Then the following limit exists and we have the equality
$$
\underset{R\rightarrow\infty}{\lim}\int\limits_{|x|\leq R}dx\,K[T](x,x)=\Tr T.
$$
\end{proposition}
Let us discuss now the case $d\geq4$. We come back to formula (\ref{B-f-calc}) and use Proposition \ref{h-dev-resolvent} with the above observations.
$$
\widetilde{g(F)}^B_\h(X) = -\pi^{-1} \lim_{\epsilon\ar 0^+} \int \!\!\int_{|\mu|\geq \epsilon}
\frac{\pa {\tilde g}}{\pa {\bar z}}(\lambda,\mu)\; r^B_{\lambda+i\mu}(X) d\lambda\,d\mu$$
\begin{equation}\label{f-calc-dev}
=-\pi^{-1} \lim_{\epsilon\ar 0^+} \int \!\!\int_{|\mu|\geq \epsilon}
\frac{\pa {\tilde g}}{\pa {\bar z}}(\lambda,\mu)\; \left[\sum_{0\leq j\leq N-1}\h^j\mathfrak{r}^B_j(\lambda+i\mu;X)+\h^{N}\widetilde{\mathfrak{r}^B_N}(\lambda+i\mu;\h,X)\right].
\end{equation}
Taking now $N>d$ and taking into account Proposition \ref{h-dev-resolvent} we conclude that $\widetilde{\mathfrak{r}^B_N}(\lambda+i\mu;\h)\in S^{(0,-N)}_1$ and by the Riemann-Lebesgue Lemma it has a continuous Fourier transform (with respect to the $\xi$ variable).
Let us consider the main terms in the expansion (\ref{f-calc-dev}) for $N>d$. Taking into account Proposition \ref{str-r-j} we have to study integrals of  the form
$$
\lim_{\epsilon\ar 0^+} \int \!\!\int_{|\mu|\geq \epsilon}
\frac{\pa {\tilde g}}{\pa {\bar z}}(\lambda,\mu)\;\mathfrak{f}^B_p(x,\xi)p_{\lambda+i\mu}^{-2-p}(x,\xi)=\mathfrak{f}^B_p(x,\xi)\frac{1}{(2+p-1)!}\lim_{\epsilon\ar 0^+} \int \!\!\int_{|\mu|\geq \epsilon}
\frac{\pa {\tilde g}}{\pa {\bar z}}(\lambda,\mu)\big(\partial_\lambda^{p+1}p_{\lambda+i\mu}^{-1}\big)(x,\xi)$$
$$
=\frac{(-1)^{p+1}}{(2+p-1)!}\mathfrak{f}^B_p(x,\xi)\big[\big(\partial_\lambda^{p+1}g\big)\circ F\big](x,\xi).
$$
We can evidently use again the Riemann-Lebesgue Lemma to obtain continuity of the Fourier transform (with respect to the $\xi$ variable).
Putting all these results together we obtain the following statement
\begin{proposition}\label{Trace-formula}~\\
For $B$ a magnetic field with components of class $BC^\infty(\mathcal{X})$ and $H$ the self-adjoint operator defined in Proposition~\ref{Prop-P-A-h}, if $g\in C^\infty_0(\mathbb{R})$ has compact support $\Sigma_g\subset(-\infty,\mathfrak{a}_0)$, then $g(H)$ is a trace-class operator\footnote{In fact it is even finite-rank.} and we have the formula
$$
\Tr g(H)=\int_{\mathcal{X}}dx\,\mathfrak{F}^-\left[\widetilde{g(F)}^B_\h\right]\left(x,0\right)=\int_{\Xi}dX\,\widetilde{g(F)}^B_\h(x,\xi),
$$
so that $\Tr g(H)$ only depends on the magnetic field.
\end{proposition}

\subsection{End of the proof of the semiclassical trace formula}

\subsubsection{Comparison of the theorems using Agmon estimates}

We are now ready to consider the semiclassical expansion of the trace
formula starting from Proposition \ref{Trace-formula} and using the
semiclassical expansions computed previously. Before doing that let us
come back more in detail  at the remark in \cite{HMR} that due to the
exponential decay of the eigenfunctions (Agmon estimates \cite{Ag})
one can modify the potential outside a compact region by polynomially
bounded terms with only an exponentially small change (of order
$\exp\{-c/\h\}$) in the eigenvalues situated in any compact part of
the discrete spectrum. A simple inspection of the proof in \cite{Ag}
shows that the same exponential decay estimates can be obtained for
the magnetic Schr\"odinger operator so that we can apply the same arguments from
\cite{HMR} to our 'magnetic' situation. Here is the basic proposition.
\begin{proposition}\label{propcomp}~\\
Let $(A,V)$ and $(\widehat A, \widehat V)$ two pairs of
electro-magnetic potentials
satisfying
 Hypotheses \ref{Hyp0} and \ref{Hyp}. Let $E$ verify
\begin{itemize}
\item $
 E < \min ( \Sigma_V,\Sigma_{\widehat V})$\,,
\item 
$U_E:= V^{-1}(]-\infty, E[)= {\widehat V} ^{-1}(]-\infty, E[)\,,$ 
\item 
$V=\widehat V$ on $U_E$ and $A= \widehat A$ on $U_E$\,,
\end{itemize}
and let $H$ and $\widehat H$ the corresponding magnetic Schr\"odinger
operators. 
Then for any $g\in C_0^\infty(\mathbb
 R)$, such that 
$
\supp g \subset ]-\infty,E[\,,
$
then $\Tr g(H)$ and $\Tr g(\widehat H)$ have the same semiclassical
expansion modulo $\mathcal O(\h^\infty)$.
\end{proposition}
It is enough to observe that, for any $\epsilon >0$, the
eigenfunctions corresponding to eigenvalues of $H$ (resp. $\widehat
H$)
 less than $E-\epsilon$ ) decay exponentially in any compact outside
 of $U_{E-\frac \epsilon 2}$. \\

This can be used in the following way.
\begin{proposition}~\\
Let $(A,V)$ satisfy Hypotheses \ref{Hyp0} and \ref{Hyp} and let
 $E <\Sigma_V$, then there exists a pair $(\widehat A,\widehat V)$
 such that the assumptions of Proposition \ref{propcomp} are satisfied
 with in addition $\widehat A$ and $\widehat V$ bounded (with all the derivatives).
\end{proposition}
The proof is easy. We can indeed consider a $C^\infty$ increasing
function
 $\chi$ on $\mathbb R$ such that
$$
\chi(t)=t \mbox{ on } ]-\infty, \frac 12 (E + \Sigma_V)[\,,\,
\chi'(t)=0 \mbox{ on } ] \frac 13 (E + 2 \Sigma_V), +\infty[\,.
$$
We can then take $\widehat V = \chi (V)$. It is not difficult to
modify
$A$ outside $V^{-1}(]-\infty,\frac 12 (E + \Sigma_V)[)$ to get
 a $C^\infty$ bounded magnetic potential.\\ 

As a consequence, it is enough for proving Theorem \ref{Semicl-trace-formula} to prove it
with $A$ and $V$ of class $C^\infty$ and bounded. Hence we can work at the
intersection
 of the two calculi and use either the results of the Weyl's calculus
 or of the adapted magnetic calculus.

\subsubsection{The case with boundary}
Let us consider the case of the Dirichlet realization in a bounded
open set $\Omega$, then it is easy to modify the comparison argument
of the previous subsubsection in order to obtain the following theorem.

\begin{theorem}\label{Semicl-trace-formulabdy}~\\
Let $A$ and $V$ be $C^\infty$ potentials on $\overline{\Omega}$ and
assume that
$$
\inf_{x\in \overline{\Omega}} V(x) < \inf_{x\in \pa \Omega} V(x)\,.
$$
Then, with
 $H$ the Dirichlet realization of $P_A$ in $\Omega$, there exists a sequence of distributions $T^B_j\in\mathcal{D}^\prime(\mathbb{R})$, ($j\in \mathbb N$),
 such that, for any $\epsilon >0$,  for any $N\in \mathbb N$, there exists $C_N$ and $h_N$,
 such that
 if  
$$ g\in C_0^{\infty}(\mathbb R)\,,\, \mbox{ with } \supp g \subset
]-\infty, \inf_{x\in \pa \Omega} V-\epsilon[\,,
$$ then, :
\begin{equation}
\left|(2\pi \h)^d\Tr\,g(H)\,-\,\sum_{0\leq j\leq
  N}\h^jT^B_j(g)\right|\,\leq\,C_N\h^{N+1}\,,\, \forall \h\in
]0,h_N]\cap \I\,.
\end{equation}
More precisely there exists
$k_j\in \mathbb N$ and universal polynomials  $P_\ell( u_\alpha, v_{\beta,j,k})$ depending on a finite
 number of variables, indexed by $\alpha \in \mathbb N^{2d}$ and
 $\beta\in \mathbb N^d$, such that the distributions:
\begin{equation}
T_j^B (g) = \sum_{0 \leq \ell \leq k_j} \int g ^{(\ell)} (F(x,\xi))
 P_\ell (\pa_{x,\xi}^{\alpha} F(x,\xi), \pa_x^{\beta} B_{jk} (x))\, dx
 d\xi\,,
\end{equation}
Finally,  $T_j^B =0$ for $j$ odd. 
\end{theorem}
\begin{remark}
 The polynomials are the same as in Theorem \ref{Semicl-trace-formula}. In particular they are independent of $\Omega$.
\end{remark}

Using a (small extension of) the comparison proposition, one can
 modify the potentials in the neighborhood of $\pa \Omega$
 and then extend outside of $\Omega$ without modify the asymptotic
 of $\Tr g(H)$ and then use the results obtained in the case of
 $\mathbb R^d$.

\begin{remark}
Note that we have not done any assumptions on the topology of
$\Omega$. Hence we have also that this expansion depends only on the
magnetic field for cases where one can get various generating magnetic
potentials which are not in the same cohomology class.
\end{remark}

\subsubsection{The odd coefficients vanish.}
To prove this result, 
one first observes that we have 
$$
\left||\h| ^d\Tr\,g(H)\,-\,\sum_{0\leq j\leq N}\h^jT^B_j(g)\right|\,\leq\,C_N\h^{N+1}\,,\, \forall \h\in [-h_N,h_N]\setminus \{0\}\,.
$$
(the $\h$-pseudodifferential calculus can be extended to $\hbar <0$)
 and using the complex conjugation one obtains that the trace of $g(H)$ is unchanged
 when $\h\mapsto -\h$. Hence the odd coefficients are $0$.

\end{document}